\documentclass{aa}
\pdfoutput=1

\usepackage{graphicx}
\graphicspath{{Images_Article/}}
\usepackage{txfonts}
\usepackage[T1]{fontenc}
\usepackage{aas_macros}
\usepackage{natbib,amsmath}

\newcommand*\unit[1]{\ensuremath{\:\mathrm{#1}}}
\newcommand*\Eth{\ensuremath{{E_\text{th}}}}
\newcommand*\Ethb{\ensuremath{{E_\text{th}^*}}}
\newcommand*\Erad{\ensuremath{{E_\text{rad}}}}
\newcommand*\Eradb{\ensuremath{{E_\text{rad}^*}}}
\newcommand*\Econd{\ensuremath{{E_\text{cond}}}}
\newcommand*\Econdb{\ensuremath{{E_\text{cond}^*}}}
\newcommand*\Amax{\ensuremath{{A_\text{max}}}}
\newcommand*\Itot{\ensuremath{I_\text{tot}}}
\newcommand*\Imax{\ensuremath{I_\text{max}}}

\begin{document}

   \title{Energetic characterisation and statistics\\ of solar coronal brightenings}

   \author{V. Joulin\inst{1},
           E. Buchlin\inst{1}\fnmsep\thanks{\email{eric.buchlin@ias.u-psud.fr}},
           J. Solomon\inst{1}
           \and
           C. Guennou\inst{2}\fnmsep\thanks{Now at LESIA, Observatoire de Paris, PSL Research University, CNRS, Sorbonne Universités, UPMC Univ. Paris 6, Univ. Paris Diderot, Sorbonne Paris Cité, 5 place Jules Janssen, 92195, Meudon, France}
          }
   \institute{Institut d'Astrophysique Spatiale, CNRS, Univ. Paris-Sud, Université Paris-Saclay, Bât. 121, 91405 Orsay, France
    \and
    Instituto de Astrof\'isica de Canarias, 38205 La Laguna, Tenerife, Spain
   }
   \date{Received ; accepted }

\abstract%
   {To explain the high temperature of the corona, much attention has been paid to the distribution of energy in dissipation events. Indeed, if the event energy distribution is steep enough, the smallest, unobservable events could be the largest contributors to the total energy dissipation in the corona. Previous observations have shown a wide distribution of energies but remain inconclusive about the precise slope. Furthermore, these results rely on a very crude estimate of the energy. On the other hand, more detailed spectroscopic studies of structures such as coronal bright points do not provide enough statistical information to derive their total contribution to heating.}
   {We aim at getting a better estimate of the distributions of the energy dissipated in coronal heating events using high-resolution, multi-channel Extreme Ultra-Violet (EUV) data.}
   {To estimate the energies corresponding to heating events and deduce their distribution, we detect brightenings in five EUV channels of the Atmospheric Imaging Assembly (AIA) on-board the Solar Dynamics Observatory (SDO). We combine the results of these detections and we use maps of temperature and emission measure derived from the same observations to compute the energies.}
   {We obtain distributions of areas, durations, intensities, and energies (thermal, radiative, and conductive) of events. These distributions are power-laws, and we find also power-law correlations between event parameters.}
  {The energy distributions indicate that the energy from a population of events like the ones we detect represents a small contribution to the total coronal heating, even when extrapolating to smaller scales. The main explanations for this are how heating events can be extracted from observational data, and the incomplete knowledge of the thermal structure and processes in the coronal plasma attainable from available observations.}


   \maketitle


\section{Introduction}
\label{sec:intro}

Following the identification of several solar coronal lines as forbidden transitions of highly-ionized atoms by Grotrian and Edlén, it was induced that the temperature of the coronal plasma reaches more than a million kelvins \citep{grotrian39, alf41, edlen43, alf47}. Seventy years after this discovery, the heating processes responsible for keeping the plasma at this high temperature despite energy losses of $300$ to $10\,000\unit{W\,m^{-2}}$ \citep{Withbroe_77} still remain largely unknown.
As large-scale, observable heating events in the solar corona do not provide sufficient power, it seems likely that the missing energy would be provided by small, unobservable events, as proposed by \citep{Parker_88} who first coined the term ``nanoflare''.
These small heating events would correspond to the dissipation of small current sheets formed in a coronal loop by the motion of magnetic field lines driven by the convective motion of the photosphere \citep{parker83}.
In such models, the magnetic energy is slowly accumulated to a ``critical'' state until it is released, in a way that can be described by Self-Organized Criticality (SOC) models \citep[e.g.][]{bak88,luh91,buchlin03}.

An alternative possibility is that the critical state is never reached so the heating is continuous.
\Citet{vanballegooijen86} obtain such a behaviour with a magnetic energy cascade model.
However, a turbulent cascade can be intermittent, and heating ``events'' can be extracted from the turbulent energy dissipation as a function of position and time \citep{einaudi96,georgoulis98}, with the caveats discussed in \citet{buchlin05}.

In both cases, once heating events and their energies are defined, the distribution of these energies can be derived.
Observationally, this distribution has in general been found to be a power-law (occurrence frequency proportional to $E^{-\alpha}$ for event energy $E$, with $\alpha$ being positive), which is consistent with SOC, turbulence, and many other mechanisms \citep{sorn}; we call $\alpha$ the ``slope'' of the power-law.
\citet{Hudson_91} noticed that small events provide the most significant contribution to the total heating only if the slope of such a power-law is steeper than $2$. However, distributions from observed flare energies in X-rays rather display a flatter slope \citep{Drake_71,lin84,dennis85,crosby93}.
For \citet{Hudson_91}, it follows that the hypothetical nanoflares must constitute a separate population of events, due to different physical mechanisms, with an energy distribution displaying a steeper slope than larger events.

This raised considerable interest in the observational determination of the slope of the heating events energy distribution, especially for smaller events, which have been observed in the Extreme Ultra-Violet (EUV) with SoHO/EIT and TRACE \citep[e.g.,][]{Berghmans_98,Krucker_98,Aschwanden_00,parnell00}.
Put together with X-ray observations of flares, these distributions extend over 8 decades in energy, with an overall power-law index of about 1.8 \citep{Aschwanden_00}.
More recently, observations with RHESSI have shown a power-law distribution with slope $1.7$ for the peak dissipation power in microflares \citep{hannah08b}, while \citet{aschwanden13b} have found a slope $1.66$ for the thermal energy computed from SDO/AIA for a selection of 155 M- and X-class GOES flares.
Nanoflare-like events had already been observed in X-rays by \citet{koutchmy97a} in coronal holes, but the number of events was too small to derive an energy distribution.

Until now and for the foreseeable future, the spatial resolution of instruments observing the corona is several orders of magnitudes larger ($100\unit{km/pixel}$ at best for the Hi-C sounding rocket, $440\unit{km/pixel}$ for SDO/AIA) than the smallest dissipation scales expected from coronal plasma properties, which could be as small as a few metres (the ion gyroradius): even the nanoflare-like brightenings in cool transition region loops seen with Hi-C \citep{winebarger13a} are still large compared to the smallest expected scales.
However, the existence of heating events at scales smaller than the observable scales is also supported by observations of their spectroscopic signatures \citep{brosius14a}.
Furthermore, the dissipation scales do not necessarily need to be observed if the statistical properties of events at these scales can be extrapolated from the statistics at somewhat larger, observable scales.
For this reason, we believe that SDO/AIA, which is the highest-resolution EUV imager available (excluding short-duration rocket flights), can provide relevant data concerning the issue of nanoflare heating of the solar corona, although it does of course not resolve dissipation scales.

Compared to what can be achieved with SDO/AIA, results based on past EUV observations could only rely on crude estimates of the energy.
For example, \citet{Berghmans_98} use two SOHO/EIT channels ($195$ and $304\unit{\AA}$) independently, and evaluate the radiative losses in each channel from the increase of emission measure, assumed to be proportional to the intensity in this channel.
They have found a power law distribution of energy with a slope $1.9$.
On other hand, \citet{Krucker_98} estimate the thermal energy from the density of the plasma, evaluated from the square root of the emission measure computed from both the $171$ and $195\unit{\AA}$ SOHO/EIT channels, and have found a power law distribution of energy with a slope $2.5$.
\citet{aletti00} have shown that part of the discrepancy between these slopes is explained by the different computation of energy, the second being proportional to the square root of the first.
\citet{aletti00} also warned that if the event volume (taking into account some filling factor) is actually dependent on event energy (while \citet{Krucker_98} assume that it is constant), one can expect a flattening of the energy distribution compared to the \citet{Krucker_98} results.

Intermediate-scale brightenings can be identified to coronal bright points, which are observed as brightenings in X-rays and EUV and correspond over their lifetime to small magnetic dipoles \citep{Madjarska_03}.
Spectroscopic studies have provided detailed physical properties of these bright points; for example, their velocity and intensity variations suggest that the heating making their plasma visible in EUV originates from the dissipation of Alfvén waves \citep{Madjarska_02}.

However, statistical studies of such brightening events are required to determine the total contribution of the corresponding heating events to coronal heating.
This is the aim of this paper, in which we use high-resolution, multi-channel SDO/AIA observations to derive as good as possible estimates of parameters of EUV brightenings, that we identify to coronal heating events.
We define EUV brightenings observationally from local intensity variations, with no a priori on what kind of structure emits EUV light.
In the next section, we present the data we use, the event detection method and the computation of event parameters (including their energy).
Then in Sec.~\ref{sec:Results} we present statistical results for event parameters distributions and correlations, and an estimate of the total energy in events, before discussing these results.

\section{Method}
\label{sec:method}

\subsection{Data}
\label{sec:data}

We use images from the Atmospheric Imaging Assembly \citep[AIA,][]{lemen12a} instrument on board the Solar Dynamics Observatory \citep[SDO,][]{pesnell12a} satellite, in 5 of the 6 coronal wavelengths observed by AIA: $13.1$, $17.1$, $19.3$, $21.1$, and $33.5\unit{nm}$.
We do not use the $9.4\unit{nm}$ channel because the signal to noise ratio is too low in Quiet Sun regions.
These images of $4096\times4096$ pixels have a pixel size of $0.6\unit{arcsec}$ ($435\unit{km}$ at Sun centre, corresponding to an area of approximately $0.19\unit{Mm^2}$), and we read them directly from the local Data Records Management System (DRMS) at the Multi-Experiment Data and Operations Centre (MEDOC) at IAS.
We select two observation periods shown in Table~\ref{tab:obs}, the first one corresponding to only Quiet Sun in the field of view, while the second one includes parts of active regions.
Both observation periods have a duration of $24$ hours and a cadence of $2$ minutes (about $720$ images each); this cadence was chosen so that Differential Emission Measure (DEM) inversions are available at the same cadence \citep{guennou12b}, allowing the computation of event energies (Sec.~\ref{sec:energy}); at higher cadence, these DEM inversions would not allow us to compute reliable energies.

\begin{table*}
  \centering
  \caption{Observing periods analysed in this paper.}
  \begin{tabular}{lllll}
    \hline\hline
    Index & Period begin & Duration & Cadence & Main solar features \\ \hline
    1 & 2011-02-08T00:00:00UT & $24\unit{hr}$ & $2\unit{min}$ & Quiet Sun \\
    2 & 2011-11-07T20:00:00UT & $24\unit{hr}$ & $2\unit{min}$ & Active Sun \\
    \hline
  \end{tabular}
  \label{tab:obs}
\end{table*}

\begin{figure*}
  \includegraphics[width=.5\linewidth]{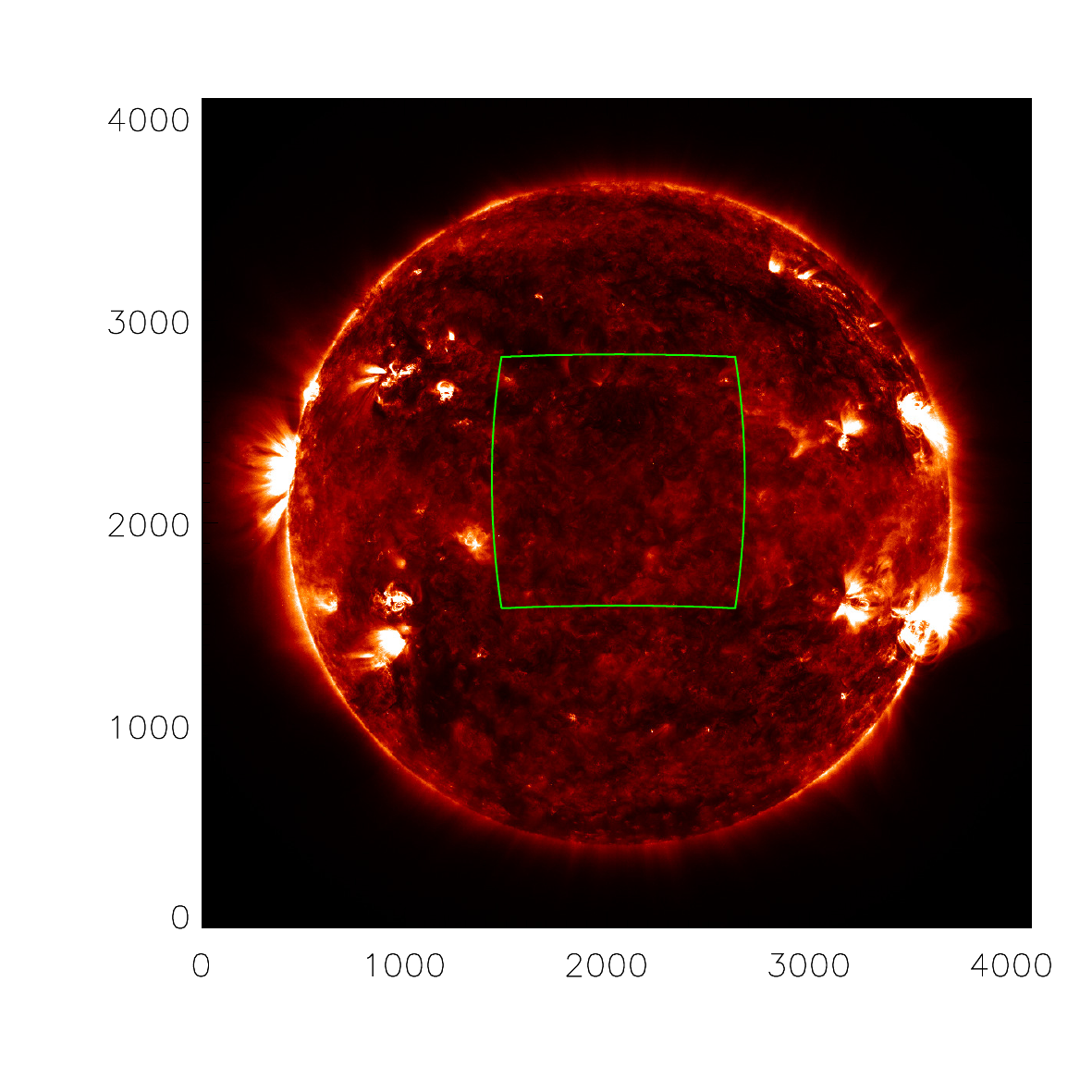}%
  \includegraphics[width=.5\linewidth]{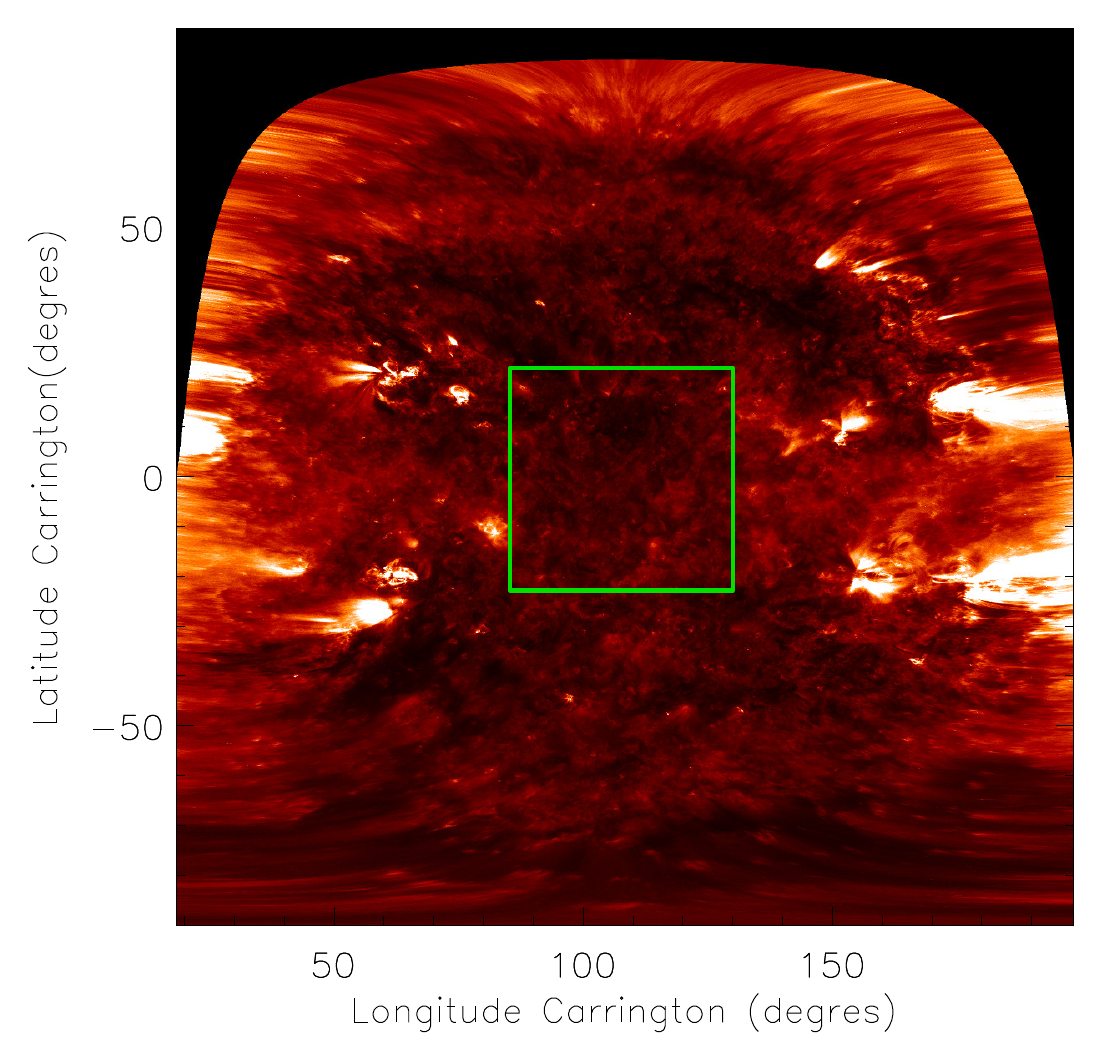}
  
  \caption{Left: SDO/AIA image in the $17.1\unit{nm}$ channel on 2011-02-08T12:00UT, with axes coordinates in pixels.
    Right: intensity map in Carrington coordinates corresponding to this SDO/AIA image.
    The selected $45 \degr \times\ 45 \degr$ region of interest is marked with a green box on both images.
  }
  \label{fig:aia_carrington}
\end{figure*}

\newcommand*\tr{\ensuremath{t_\textrm{ref}}}
In order to follow coronal features with solar rotation, we work in Carrington heliographic coordinates \citep{thompson06}, compensated for the effect of solar differential rotation between a reference time \tr\ (chosen as being the middle of the observing period) and the time $t_k$ for each image of index $k$.
We select a $\Delta \phi \times \Delta \theta = 45 \degr \times\ 45 \degr$ region of interest which is centred on the solar equator and central meridian at the reference time, i.e. on a heliographic latitude $\theta=0 \degr$ and a longitude $\phi$ equal to the Carrington longitude of the observer $\phi_{obs}$ as given by the \verb+CRLN_OBS+ FITS keyword of SDO/AIA at time \tr.
This window is fixed in the compensated Carrington coordinates, ensuring that features remain at a fixed position  (unless they have some proper motion) and do not move out of the region of interest during the observing duration.

Starting from pixel $(i_C,j_C,k)$ in image $k$ in compensated Carrington coordinates, we obtain successively \citep{thompson06}:
\begin{itemize}
  \item The Carrington coordinates, compensated for differential rotation:
  \begin{equation}
  \label{eq:carrington}
    \begin{cases}
      \phi_0 &= i_C \, \delta\phi + (\phi_{\text{obs}}(\tr) - \Delta\phi/2) \\
      \theta_0 &= j_C \, \delta\theta - \Delta\theta/2
    \end{cases}
  \end{equation}
  where $\delta\phi$ and $\delta\theta$ are the chosen pixel sizes in longitude and latitude.
  
  \item The Stonyhurst coordinates:
  \begin{equation}
    \label{eq:stonyhurst}
    \begin{cases}
      \phi &= \phi_0 + (t - t_\text{ref}) (a+b\,\sin^2 \theta_0 + c\,\sin^4\theta_0) - \phi_{\text{obs}}\\
      \theta &= \theta_0
    \end{cases}
  \end{equation}
  where $\phi_{\text{obs}}$ is the Carrington longitude of the observer at reference time $t_\text{ref}$, and $(a,b,c)$ are the solid and differential rotation coefficients from small UV structures by \citet{hortin}.
  
  \item The corresponding Cartesian coordinates on a sphere of radius 1:
  \begin{equation}
    \label{eq:cartesian}
    \begin{cases}
      x_1 &= \cos \theta \, \sin \phi \\
      y_1 &= \sin \theta \\
      z_1 &= \cos \theta \, \cos \phi 
    \end{cases}
  \end{equation}
  
  \item The coordinate system is then rotated with respect to the $x_1$ axis by the Carrington latitude of the observer $\theta_
  \text{obs}$ (FITS keyword \verb+CRLT_OBS+) so that the new $z$ axis is towards the observer:
  \begin{equation}
    \label{eq:cartesian2}
    \begin{cases}
      x_2 &= x_1 \\
      y_2 &= y_1 \, \cos \theta_\text{obs} - z_1 \, \sin \theta_\text{obs} \\
      z_2 &= z_1 \, \cos \theta_\text{obs} + y_1 \, \sin \theta_\text{obs}
    \end{cases}
  \end{equation}
  The $z$ coordinate (orthogonal to the plane of sky) can then be discarded.

  \item The instrument roll angle $\alpha$ is taken into account by a rotation of the coordinate system with respect to the $z_2$ axis:
  \begin{equation}
    \label{eq:cartesian3}
    \begin{cases}
      x_3 &= x_2 \, \cos \alpha - y_2 \, \sin \alpha \\
      y_3 &= x_2 \, \sin \alpha + y_2 \, \cos \alpha
    \end{cases}
  \end{equation}
  
  \item We finally obtain coordinates in pixels on the SDO/AIA image by:
  \begin{equation}
    \label{eq:image}
    \begin{cases}
      i &= R_\sun \, x_3 + i_r \\
      j &= R_\sun \, y_3 + j_r
    \end{cases}
  \end{equation}
  where $R_\sun$ is the solar radius in pixels (\verb+R_SUN+ FITS keyword) and $(i_r, j_r)$ is the position of disk centre on the image (the \verb+CRPIX1+ and \verb+CRPIX2+ FITS keywords, as \verb+CRVAL1+ and \verb+CRVAL2+ are zero in Level-1 SDO/AIA images).
\end{itemize}

With a pixel size $\delta\phi=\delta\theta \approx 0.0277\degr$ in Carrington longitude and latitude (about $340\unit{km}$ on the solar equator), we obtain images of $1626$ pixels $\times$ $1626$ pixels.
The intensity value of each pixel $(i_C, j_C, k)$ in these images in compensated Carrington coordinates is obtained by computing the coordinates $(i,j,k)$ of the corresponding position in the original image at time $t_k$ with Eq.~\eqref{eq:carrington}--\eqref{eq:image}, and taking the bilinear interpolation of the intensities in the four neighbouring pixels.
Images in compensated Carrington coordinates are stacked in the time direction, giving a data cube indexed by $(i_C, j_C, k)$; in the following, if not stated otherwise, all Carrington coordinates and maps are assumed to be the ones compensated for solar rotation.

\subsection{Event detection}

Intensities in the images (in SDO/AIA data units) are first normalized by the exposure time in seconds.
To correct the long-term ($>4\unit{hr}$) trend in global activity, we build the time series of the mean intensity as a function of time, we normalize it to an average value of $1$, we smooth it with a running average over $4\unit{hr}$, and we divide each image $k$ by the value of the smoothed time series at time $t_k$.
For each pixel $(i_C,j_C)$ in the Carrington map, we compute an intensity threshold $\tau(i_C,j_C)$ defined from the temporal mean $\bar  I(i_C,j_C)$ of the pixel intensity $I(i_C, j_C, k)$ over the whole observation duration and its standard deviation $\sigma(i_C, j_C)$ by $\tau = \bar I + 2 \sigma$.
The threshold is different for different pixels, and so it is possible to detect small brightenings in different regions of the Sun simultaneously even if they have different background intensities.
The coefficient 2 comes from a compromise between not missing small events, and constraints due to noise (see Sec.~\ref{sec:noise}).

Candidate events are connected clusters of pixels $(i_C,j_C,k)$ which have an intensity larger than the threshold $\tau(i_C,j_C)$.
To build these clusters, we consider as neighbours of the current pixel the 26 pixels touching its faces, edges, and corners.

\subsection{Event parameters}
\label{sec:bandparams}

\begin{figure*}[htbp]
\centering
   \includegraphics[width=\linewidth]{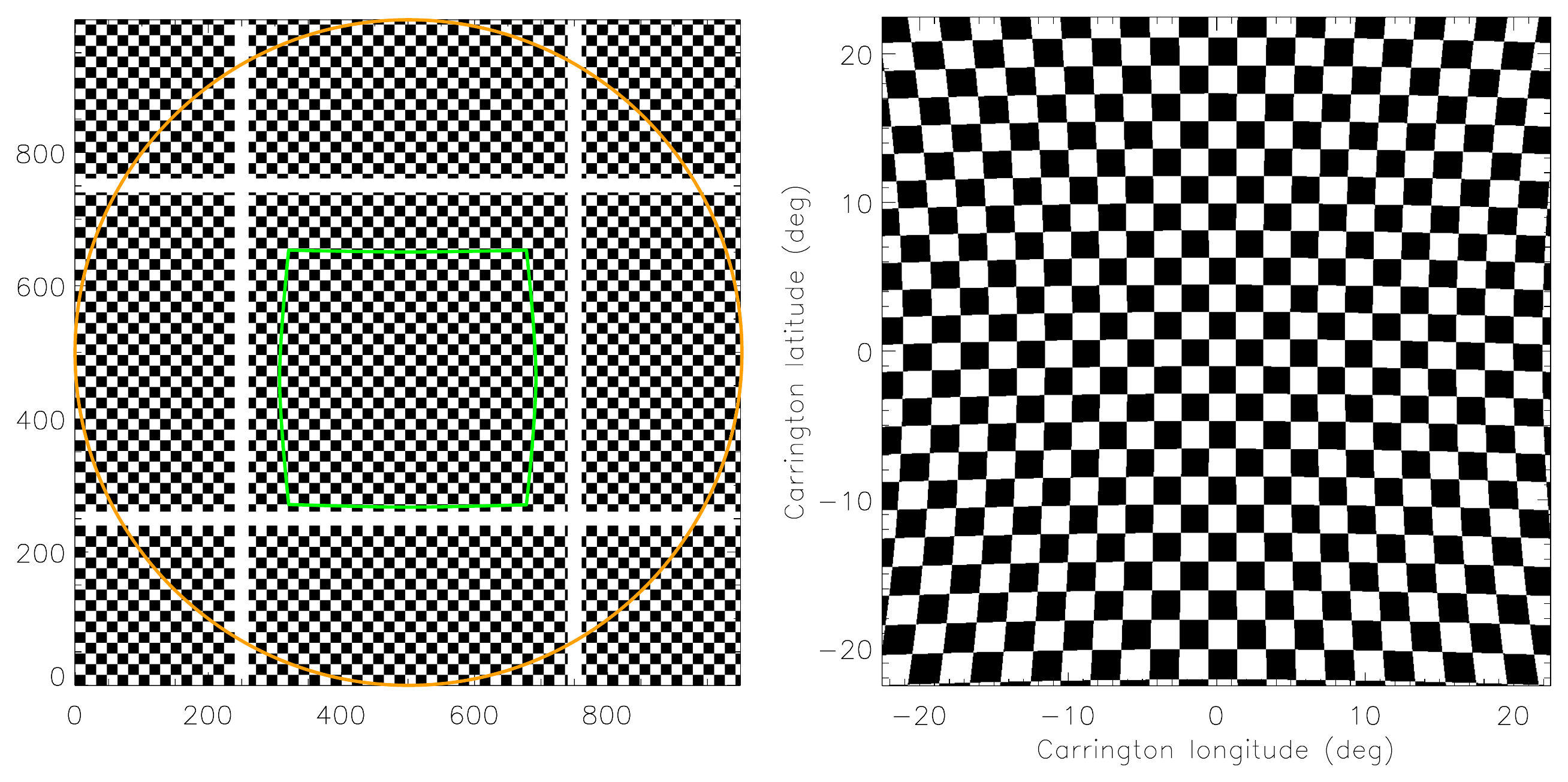}
   \caption{Left: test image with a grid pattern, corresponding to a ``Sun'' (with the limb shown in orange) with $\phi_{obs}=0\degr$, $\theta_\text{obs}=5\degr$, $\alpha=0\degr$, $R_\sun = 500$ pixels; axes are labelled in pixels. Right: the $45\degr \times 45\degr$ central field of view (shown in green on the left image) projected to a Carrington coordinates grid with a pixel size $\delta\theta = \delta\phi=0.0277\degr$.}
  \label{fig:grid}
\end{figure*}

For each event, we compute different parameters: event duration, time-dependent area and intensity, and total and maximum intensity.
We will use these parameters later on to compute the event energies.

The event start time is defined by the time of the first image (of index $k_\text{start}$) in which the event is detected, minus half the observing cadence.
Similarly, the event end time is the time of the last image (of index $k_\text{end}$) in the event, plus half the observing cadence.
The event duration $D$ is then the difference between the event end and start times.

To obtain areas as a function of time in the event, we take into account the dilation of solar structures on the edges of the field of view compared to the disk centre due to the transform to Carrington coordinates.
The dilation of SDO/AIA pixels, as well as their deformation due to the projection to a Carrington grid, can be seen close to the solar limb on Fig.~\ref{fig:aia_carrington}, and is shown more in detail in Fig.~\ref{fig:grid} for a test pattern on the $45\degr\times45\degr$ region of interest that we consider.
The dilation (or scaling) factor $d$ (number of pixels in the SDO/AIA image corresponding to one given pixel $(i_C,j_C,k)$ of the Carrington image) is the Jacobian determinant of the transform defined by the composition of the functions defined by Eqs.~\eqref{eq:carrington} to \eqref{eq:image}.
It can easily be computed from the product of the Jacobian determinants of each of these functions:
\begin{align}
  d(i_C,j_C,k) &=
  \left|
    \begin{array}{cc}
      \partial i / \partial i_C & \partial i / \partial j_C \\
      \partial j / \partial i_C & \partial j / \partial j_C
    \end{array}
  \right| \notag \\
  &= \delta\phi \, \delta\theta \, R_\sun^2
  \left(\vphantom{\cos^2}
    \cos\theta(i_C,j_C,k) \, \sin\theta(i_C,j_C,k) \, \sin\theta_\text{obs} \right.\notag \\
    & \qquad \left. + \cos\phi(i_C,j_C,k) \, \cos^2\theta(i_C,j_C,k)\, \cos\theta_\text{obs}
  \right) 
\end{align}
with $\theta(i_C,j_C,k)$ and $\phi(i_C,j_C,k)$ defined by \eqref{eq:carrington}-\eqref{eq:stonyhurst} at time $t_k$.
With our values of $\delta\phi$ and $\delta\theta$, this number is about 0.6 at disk centre, and decreases by 20\% close to the edges of the $45\degr \times 45\degr$ region of interest.

The area on the plane-of-sky (in m$^2$) corresponding to a pixel $(i_C,j_C,k)$ in the Carrington maps is then
\begin{equation}
  A_\text{pC}(i_C,j_C,k) = A_\text{pA} \, d(i_C, j_C, k)
\end{equation}
where the pC subscript denotes a quantity per pixel in the Carrington maps while the pA subscript denotes a quantity per pixel in the original SDO/AIA images; as mentioned earlier, $A_\text{pA} \approx 0.19\unit{Mm^2}$.

Note that the correction of differential rotation (a shear transform in Carrington coordinates) and the instrument roll angle (corresponding to a rotation of the image) have no impact on this dilation factor.
By comparing test areas transformed to Carrington coordinates to the theoretical areas obtained by the dilation factor computation, we have checked that errors (due to rounding errors on projected pixel coordinates) were smaller than 1\% over the field of view.

The event area at a given time $t_k$, projected on the plane-of-sky, is
\begin{equation}
  A(t_k) = \sum_{(i_C,j_C,k)\in \text{event}} \!\!\!\! A_\text{pC}(i_C,j_C,k) = A_\text{pA} \!\!\!\! \sum_{(i_C,j_C,k)\in \text{event}} \!\!\!\! d(i_C, j_C, k)
\end{equation}
where the sum is computed at given $k$, on all $(i_C,j_C)$ such that $(i_C,j_C,k)$ is in the event (the same notation will be used in Eq.~\ref{eq:itk}, \ref{eq:ethtk}-\ref{eq:ethbtk}, and \ref{eq:pradtk}-\ref{eq:pradbtk}).

As the Carrington images are obtained by interpolation of the original SDO/AIA image, we also have to take into account the area dilation factor of each pixel in Carrington coordinates to compute the time-dependent, plane-of-sky area-integrated intensity in the event for any given time $t_k$:
\begin{equation}
  \label{eq:itk}
  I(t_k) = \sum_{(i_C,j_C,k)\in \text{event}} I(i_C,j_C,k) \, d(i_C,j_C,k)
\end{equation}

The total and maximum intensity in the event are finally defined by
\begin{align}
  \Itot &= \sum_k I(t_k) & \Imax & = \max_k I(t_k)
\end{align}

\subsection{Noise estimation and detection thresholds}
\label{sec:noise}

To exclude detections due to noise, we discard events which have a volume (total number of pixels in the SDO/AIA images, that is the sum of $A(t_k) / A_\text{pA}$ over the duration of the event) of at most 9 pixels.
This threshold has been chosen because it is the volume from which the number of detections in a data cube with a Poisson noise (with the same average as the low signal-to-noise $13.1\unit{nm}$ data cube in period~1) becomes smaller than 10\% of the number of detections in the corresponding SDO/AIA data cube.
We also discard events detected on only one image, as they may be due to cosmic rays.

\subsection{Combination of events detected in the five bands}
\label{sec:combination}

As we want to obtain a unique set of events to compute their energies, we combine the events detected in different bands.
We start by selecting all pixels belonging to events detected in at least two bands (as, given the SDO/AIA temperature response functions, a brightening is not expected to be seen in one band only).
Then we define ``combined events'' as connected clusters of pixels in this new set of pixels, with the same duration and area thresholds as for event detections in individual bands.
This further reduces the probability of detecting noise.

In this way, we obtain an additional set of events, for which we compute the duration and area as a function of time, in the same way as we did in Sec.~\ref{sec:bandparams} for events detected in each band.

\subsection{Event energies}
\label{sec:energy}

\subsubsection{Thermal energy}
\label{sec:eth}

To compute the time dependent thermal energy of each combined event, we first need to estimate the temperature and density associated to each pixel.
We use maps of temperature $T$ and emission measure $EM$ derived by \citet{guennou12b} from the same SDO/AIA EUV images (in all coronal channels, including the $9.4\unit{nm}$ bandpass).

These maps are transformed to the Carrington coordinates $(i_C,j_C,k)$ defined by Eq.~\eqref{eq:carrington}-\eqref{eq:image}.
We estimate the electronic number density assuming an event line-of-sight thickness $h$ and a filling factor $q$, by $n(i_C,j_C,k)=\sqrt{EM(i_C,j_C,k)/(q\,h)}$, using the $EM$ integrated over the line-of-sight, in $\unit{m^{-5}}$.
The value of $h$ for the event cannot be known directly from the observations but can be taken as a given function of the event maximum area; we choose $q=1$ and $h = \sqrt{\Amax}$ with $\Amax = \max_k A(t_k)$, meaning that the line-of-sight thickness of the event is assumed to remain of the same order as its characteristic maximum size on the plane of sky.

Assuming a fully ionized hydrogen plasma, the thermal energy associated to one pixel $(i_C,j_C,k)$ in the event is then
\begin{align}
  E_\text{th,pC} &= 3 k_{B} T \, (A_\text{pC} \, h \, q) \, \sqrt{EM / (h \, q) } \notag \\
  &= 3 k_{B} T \, A_\text{pC} \, \sqrt{EM \, h \, q }
\end{align}
where $k_{B}$ is the Boltzmann constant and where $E_\text{th,pC}$, $T$, $A_\text{pC}$, and $EM$ are functions of $(i_C,j_C,k)$.

We use the same method to compute the thermal energy associated to background pixels, defined as the set of pixels $(i_C,j_C,k_\text{start}-1)$ such that $(i_C,j_C,k_\Amax)$ is included in the event, where $k_\Amax$ is the first time index when the event reaches its maximum area.
The background thermal energy per pixel $E^{(\text{bg})}_\text{th,pC}$ is taken as the median of the thermal energies associated to the pixels of the background.

Then the time dependent thermal energy of the event, without (\Eth) and with (\Ethb) background subtraction, is:
\begin{align}
  \label{eq:ethtk}
  \Eth(t_k) &= \sum_{(i_C, j_C, k)\in\text{event}} E_\text{th,pC}(i_C,j_C,t_k)\\
  \label{eq:ethbtk}
  \Ethb(t_k) &= \sum_{(i_C, j_C, k)\in\text{event}} \left( E_\text{th,pC}(i_C,j_C,t_k) - E^{(\text{bg})}_\text{th,pC} \right)
\end{align}

In the following, when not written explicitly otherwise, we only consider the maximum thermal energies (noted $\Eth$ and $\Ethb$ with no time dependence) in statistics, as they are the relevant quantity for the global energy balance in events: they represent the variation of the thermal energy, first an increase due to heating, then a decrease due to energy losses (in particular by radiation and conduction).

\subsubsection{Radiative energy}
\label{sec:erad}

The radiative loss power per unit volume is computed by $n^2 \Lambda(T)$ where $\Lambda(T)$ is the optically thin radiative loss function of \citet{klimchuk08a}.
The radiative loss power associated to one pixel $(i_C,j_C,k)$ in the Carrington maps is then
\begin{equation}
  \label{eq:erad}
  P_\text{rad,pC} = A_\text{pC} \, EM \, \Lambda(T)
\end{equation}
where $P$, $A_\text{pC}$, $EM$, and $T$ all depend on $(i_C,j_C,k)$.
The time dependant radiative loss power, without and with background subtraction, is then
\begin{align}
  \label{eq:pradtk}
  P_\text{rad} (t_k) &= \sum_{(i_C, j_C, k)\in\text{event}} P_\text{rad,pC}(i_C,j_C,k)\\
  \label{eq:pradbtk}
  P^*_\text{rad} (t_k) &= \sum_{(i_C, j_C, k)\in\text{event}} \left(P_\text{rad,pC}(i_C,j_C,k) - P^{(\text{bg})}_\text{rad,pC}\right)
\end{align}
where the background radiative loss power per pixel $P^{(\text{bg})}_\text{rad,pC}$ is the median of the values given by Eq.~\eqref{eq:erad} applied to the background pixels already defined for the thermal energy.
The total radiated energy $\Erad$ (or $\Eradb$ with background subtraction) in the event is the integral of this power over the lifetime of the event.

\subsubsection{Conduction energy}
\label{sec:econd}

We start with the \citet{spitzerharm53} conductive heat flux along the loop magnetic field:
\begin{equation}
  F_c = -\kappa_0 \, T^{5/2} \, \frac{\partial T}{\partial s}
\end{equation}
with $\kappa_{0} = 10^{-11} \mathrm{W\cdot m^{-1} \cdot K^{-\frac{7}{2}}}$.
In order to use the estimate derived by \citet{klimchuk08a} for their 0D loop model, we assume that an event corresponds to a cylindrical loop of half length $L = \sqrt{A_\text{max}} / 2$ and of radius $r=L/a$ with $a=10$.
We can then give a tentative estimate of the power conducted from the corona to the chromosphere as
\begin{equation}
  P_\text{cond} (t_k) = \frac27 \pi \, \kappa_0 \, \frac{L}{a^2} T^{7/2}
\end{equation}
where $T$ is the maximum observed temperature in the event at time $t_k$.
If $T_{(\text{bg})}$ is the median temperature in the background pixels defined in Sec.~\ref{sec:eth}, we can also define a background-subtracted conducted power
\begin{equation}
  P^*_\text{cond} (t_k) = \frac27 \pi \, \kappa_0 \, \frac{L}{a^2} \left (T^{7/2} - T_{(\text{bg})}^{7/2} \right)
\end{equation}
and the corresponding energies $E_\text{cond}$ and $E^*_\text{cond}$ after integration over the lifetime of the event.

\section{Results}
\label{sec:Results}

\begin{table*}
  \caption{For each of the used SDO/AIA bands: typical temperatures of high intensity contribution (bold: main contribution in the case of a Quiet Sun DEM), average intensity per pixel of the region of interest over both observing periods, and number of detected events (in each band and for combined events).}
  \label{tab:numbers}
  \centering
  \begin{tabular}{lrrrrrr}
    \hline\hline
    & 13.1 & 17.1 & 19.3 & 21.1 & 33.5 & Combined \\ \hline
    $\log T_{\textrm{max}}$ [K] & \textbf{5.6}, 7.0 & \textbf{5.8} & \textbf{6.1}, 7.3 & \textbf{6.3} & 5.9, \textbf{6.4} \\
    \multicolumn{7}{l}{Average intensity per pixel in region of interest:} \\
    ~~~~~Obs.\ period~1 & 2.0 & 94 & 67 & 15 & 0.81 \\
    ~~~~~Obs.\ period~2 & 7.3 & 216 & 285 & 109 & 8.2 \\ 
    \multicolumn{7}{l}{Number of detected events:} \\
    ~~~~~Obs.\ period~1 & 293013 & 82600 & 78358 & 139799 & 192721  &  63154 \\
    ~~~~~Obs.\ period~2 &  204135 & 66448 & 48947 & 70154 & 192475 & 47158 \\
    \hline
  \end{tabular}
\end{table*}

\begin{figure*}
  \centering
  \includegraphics[width=.5\linewidth]{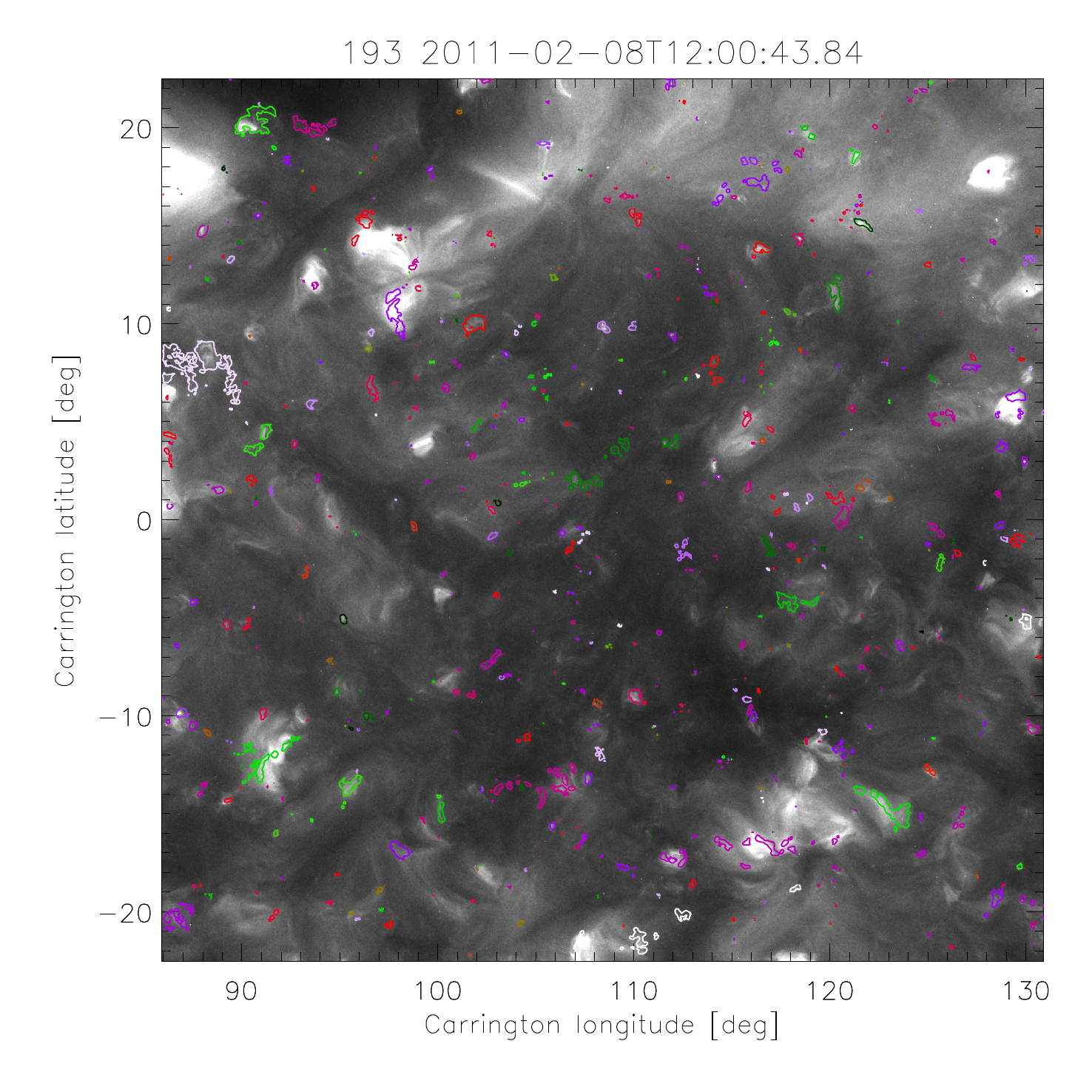}%
  \includegraphics[width=.5\linewidth]{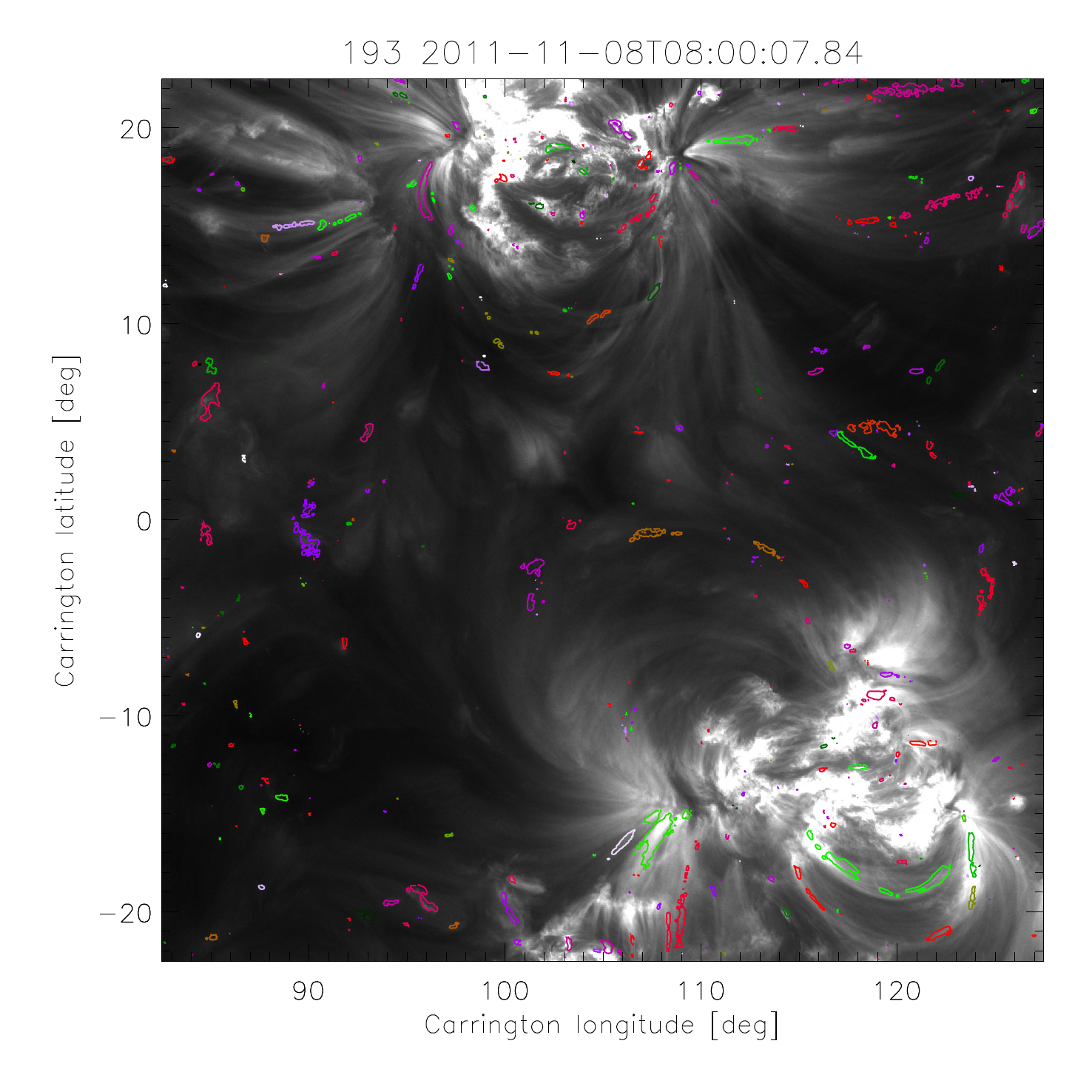}
  \caption{Cut of event boundaries at the middle of period~1 (left) and 2 (right), on the corresponding SDO/AIA $19.3\unit{nm}$ images as background, in Carrington coordinates.
  Different colours indicate different events (but different events may have the same colour).
  Movies are available with the electronic version of the journal.
  }
  \label{fig:events}
\end{figure*}

\subsection{Detected events}

The number of events detected in both observing periods for all types of events (in the five selected SDO/AIA bands as well as the ``combined'' events) is shown in Table~\ref{tab:numbers}.
Although these numbers are dependent on the detection method and thresholds used, it can be noticed that almost all types of events are significantly more numerous in period~1 (quiet Sun) than in period~2 (active Sun).
There are also more detections in the bands with the lower signal ($13.1\unit{nm}$ and $33.5\unit{nm}$), although the thresholds chosen in Sec.~\ref{sec:noise} ensure that no more than 10\% of these events are false detections due to noise.

The combined events seen at the  middle of each observing period are represented in Fig.~\ref{fig:events}, and the movies representing them over the full observing periods are available with the electronic version of the journal.
In observing period~1, there are small events quite uniformly spread over the field of view.
Some of them are correlated to bright structures like small loops that are part of coronal bright points, while others appear in quieter regions.
Conversely, not all bright structures are detected as events, as structures which remain bright over long time scales compared to the observing times are not selected by the detection algorithm.
This is because the algorithm is designed to find brightenings at small time scales compared to the observing time scale.
The same holds for period~2, in which some events are part of quiet regions, while others are correlated to the active regions' loops and moss.
Like the coronal bright points of period~1, the active regions of period~2 are not detected per se, but the brightenings occurring in them are detected.

\subsection{Event distributions}
\label{sec:distributions}

We then perform a statistical analysis of the event parameters (including the energies), that are obtained using the equations of Sec.~\ref{sec:bandparams} and~\ref{sec:energy}, starting with their distributions.
To get a frequency distribution (number of events per unit time, per unit area, and per unit of any given event parameter), we build a histogram of the event parameter with bins of constant width in logarithmic space, and we divide it by the width (in linear space) of each bin, by the observation duration, and by the region of interest area.
This area is $R_\sun^2 \, \frac\pi2 \sin(\frac\pi8)$ for the $45\degr \times 45\degr$ region of interest that we have chosen (about $1/21$ of the total solar sphere area), with $R_\sun$ given in physical units.

\paragraph{Maximum areas.}

\begin{figure*}
  \includegraphics[width=.32\linewidth]{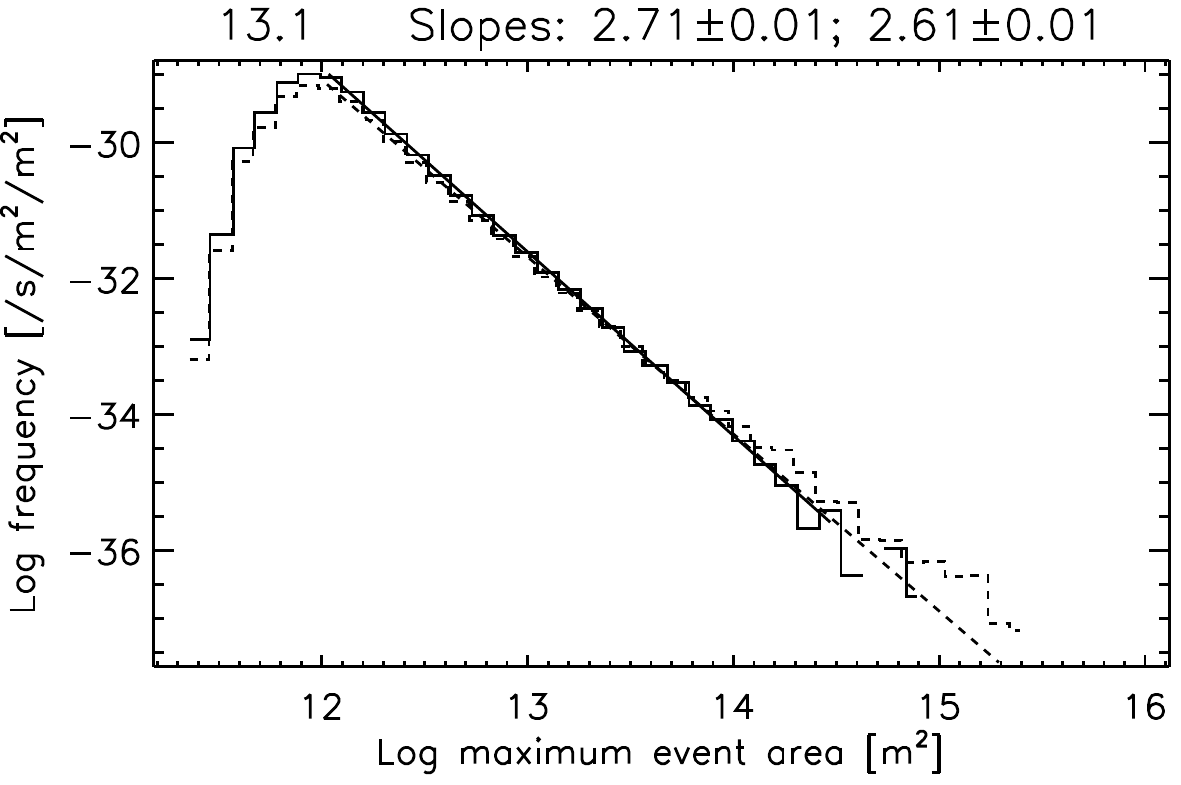}
  \includegraphics[width=.32\linewidth]{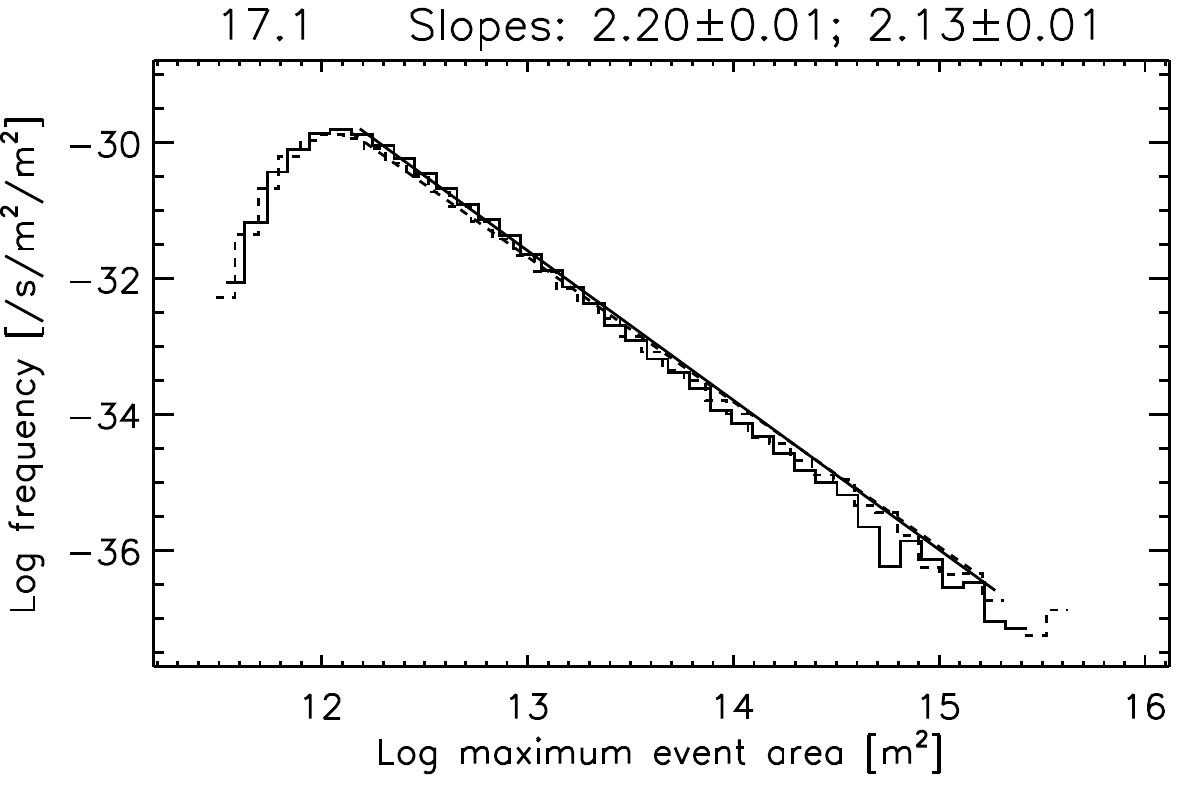}
  \includegraphics[width=.32\linewidth]{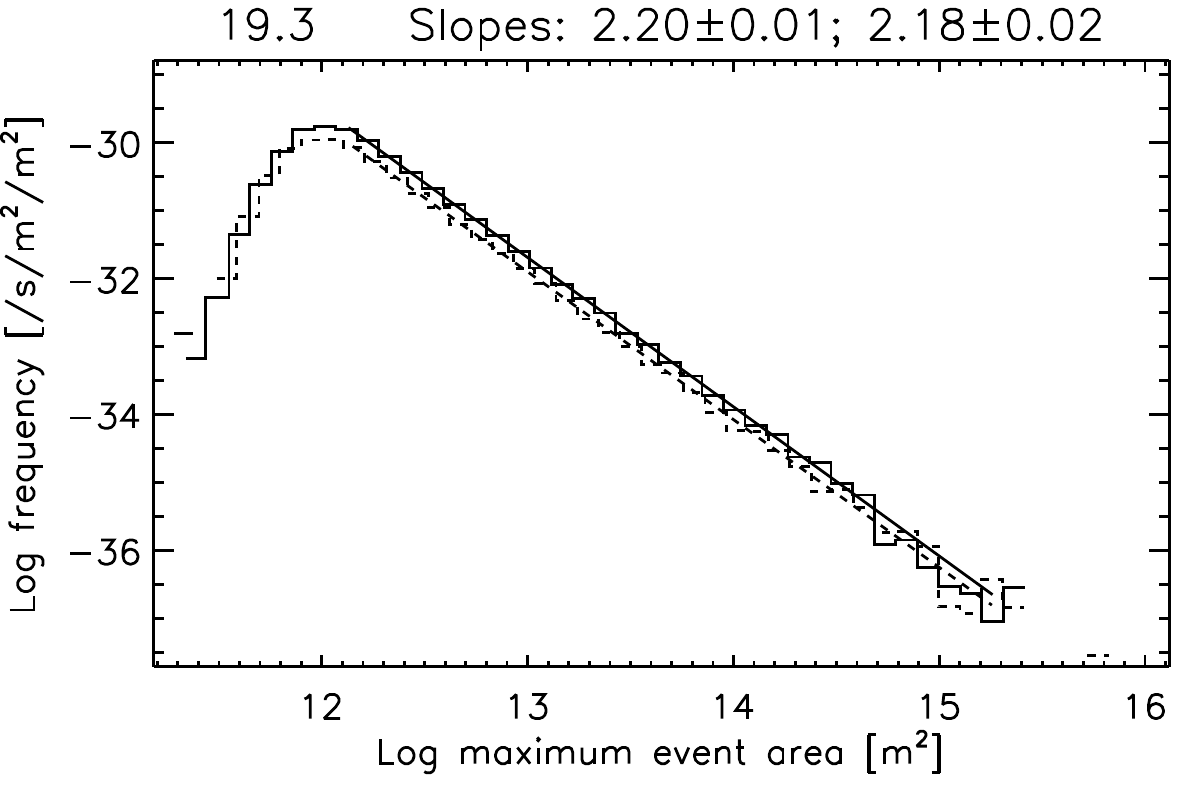} \\
  \includegraphics[width=.32\linewidth]{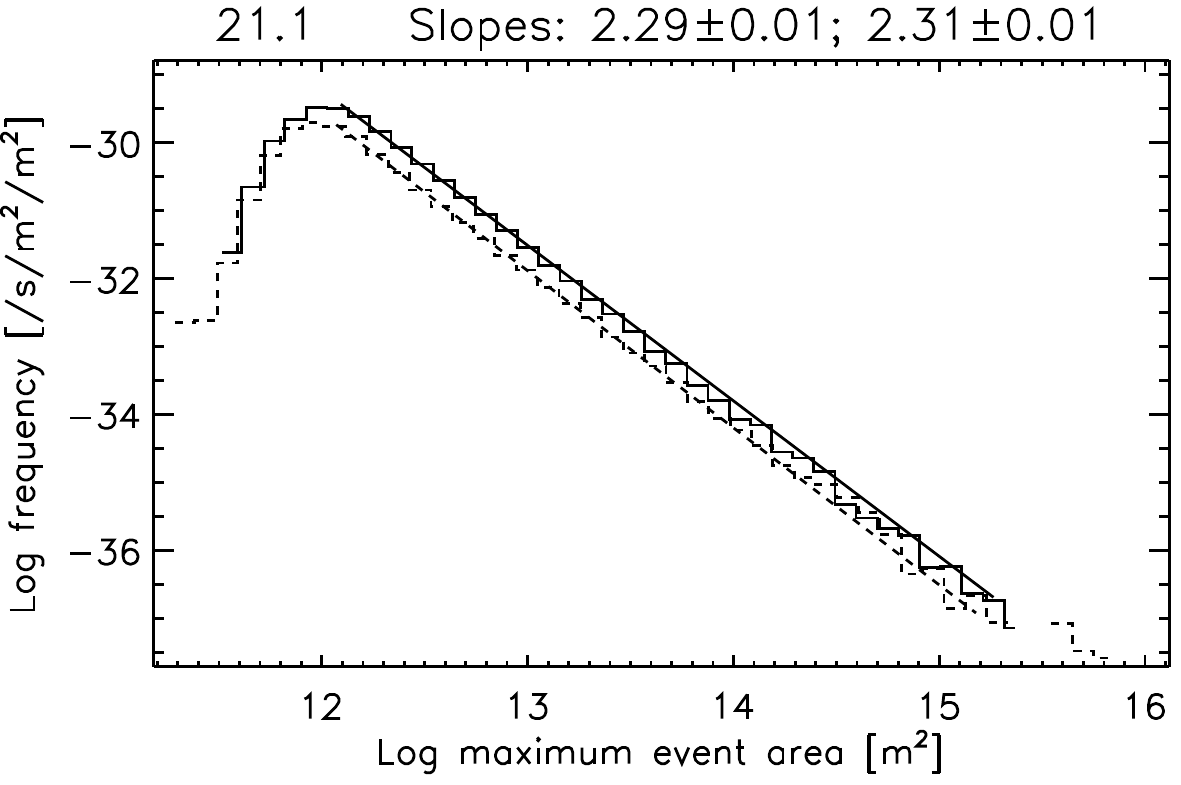}
  \includegraphics[width=.32\linewidth]{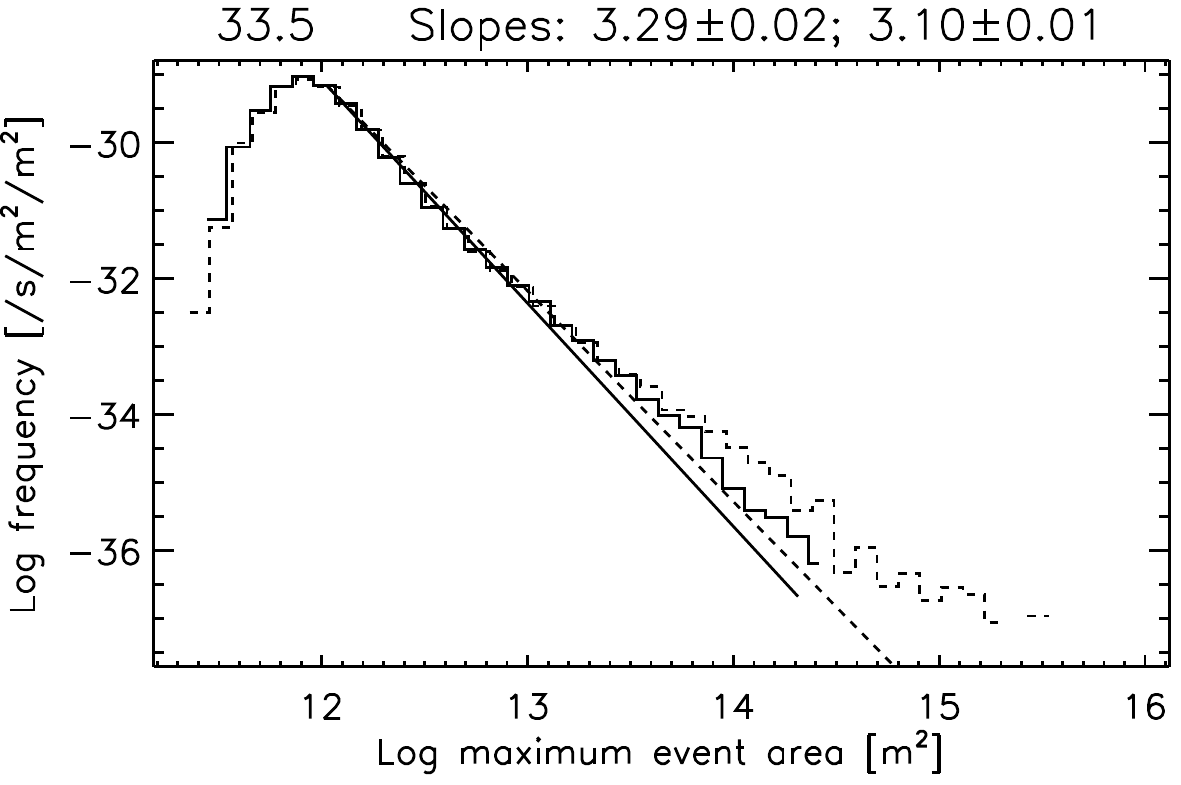}
  \includegraphics[width=.32\linewidth]{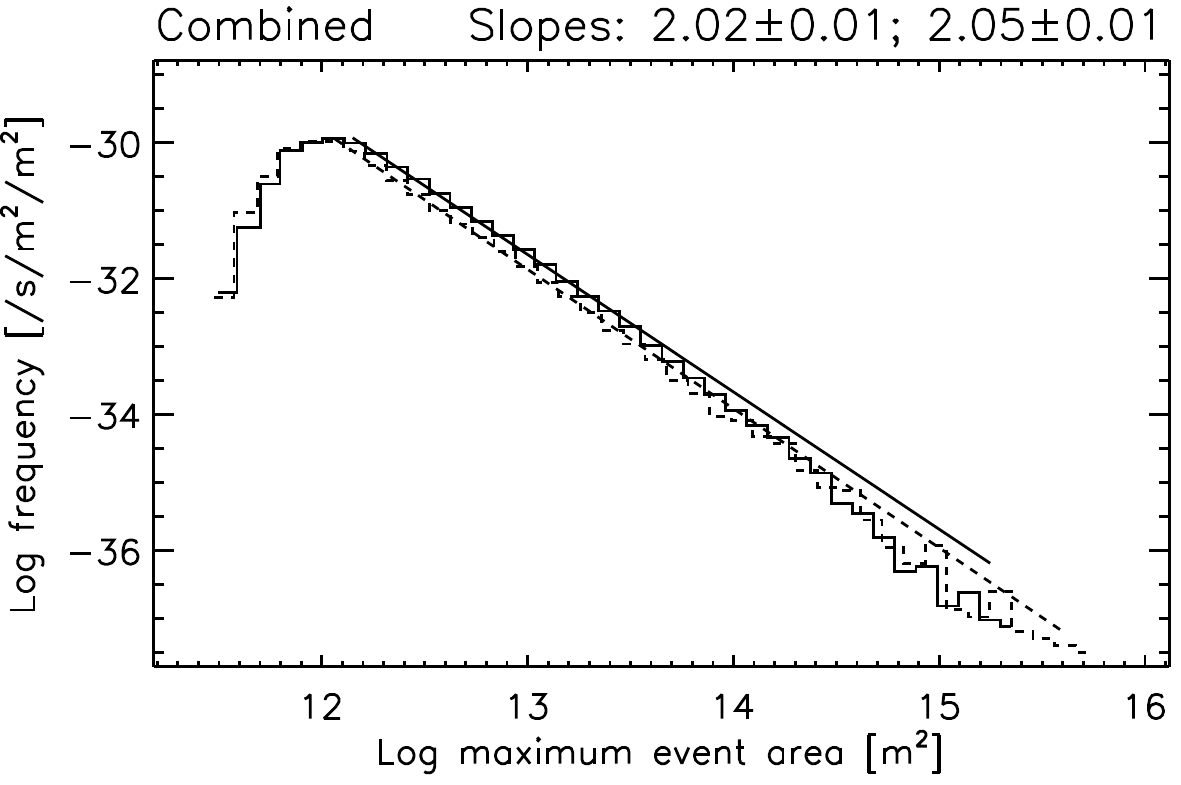}
  \caption{From left to right and from top to bottom: frequency distributions of maximum areas $A_\text{max}$ for events detected in the $13.1$, $17.1$, $19.3$, $21.1$, and $33.5\unit{nm}$ SDO/AIA bands, and for events derived from the combined events. The plain lines are for observing period~1 and the dashed lines for observing period~2. Power law fits of both distributions are shown, and their slopes are  displayed above each plot (for period~1, then for period~2).}
  \label{fig:distareas}
\end{figure*}

The frequency distributions of event maximum areas $A_\text{max}$ (Fig.~\ref{fig:distareas}) detected in the SDO/AIA bands are power-laws over about $3$ orders of magnitude, extending from $10^{12}$ to $10^{15}\unit{m^2}$ ($5$ to $5000$ SDO/AIA pixels), for both observing periods.
An exception is the reduced range, corresponding to a lack of events larger than $2\times 10^{14}\unit{m^2}$, of the area distribution for events detected in the $33.5\unit{nm}$ band in period~1: no large events are detected in this band corresponding to hotter plasma than the other bands (excluding $13.1\unit{nm}$, which also contains emission from much cooler plasma; see Table~\ref{tab:numbers}).

We perform for each frequency distribution a linear fit in logarithmic axes, giving the parameters of a power law.
The fitting range is determined automatically, when possible, as the widest range with no zero-frequency bins between the value of $A_\text{max}$ corresponding to the maximum frequency and the maximum value of $A_\text{max}$.
This fit takes into account the uncertainty in each point, computed assuming Poissonian statistics of the number of events in each of the histogram bins that were used to compute the frequency distribution.
As a consequence, the fit gives less weight to points with a small number of events per histogram bar, which is in particular the case (given the slopes that we obtain) at the rightmost end of the fitting range.
These uncertainties on the frequency distributions yield uncertainties on the fitted slopes, that are indicated on the plots with the slope values.

The power-law slopes of the fits of the maximum area distributions are then (in absolute values) between $2.13$ and $2.31$ for the $17.1$, $19.3$, and $21.1\unit{nm}$ bands in both observing periods, while they are steeper ($2.61$ and $3.10$) for the $13.1$ and $33.5\unit{nm}$ bands in period~2 (active Sun), and even steeper ($2.71$ and $3.29$) in period~1 (quiet Sun).
These steeper slopes could be interpreted as an over-representation of small events in these bands, specifically in the quiet Sun.
In the case of the $33.5\unit{nm}$ band however, in both observing periods, the slope of the distribution of the small number of large events is reduced (to about $2.5$) with respect to the overall fitted slope (which is dominated by the smaller events): there are more large events than expected from an extrapolation of the maximum area distribution for small events.

The range of maximum areas for the combined events is comparable to the ones for the events detected in the individual SDO/AIA bands, although, as it could be expected, some larger events are detected in period~2 (up to $5\times 10^{15}\unit{m^2}$) than in period~1.
Compared to the distributions for events detected in individual bands, the power-law slopes are reduced ($2.02$ and $2.05$), and the frequencies of small events ($10^{12}\unit{m^2}$) are smaller.
This suggests that some small events detected in individual bands have been either discarded or merged with events detected in other bands during the procedure leading to ``combined'' events.

\paragraph{Durations.}

\begin{figure*}
  \includegraphics[width=.32\linewidth]{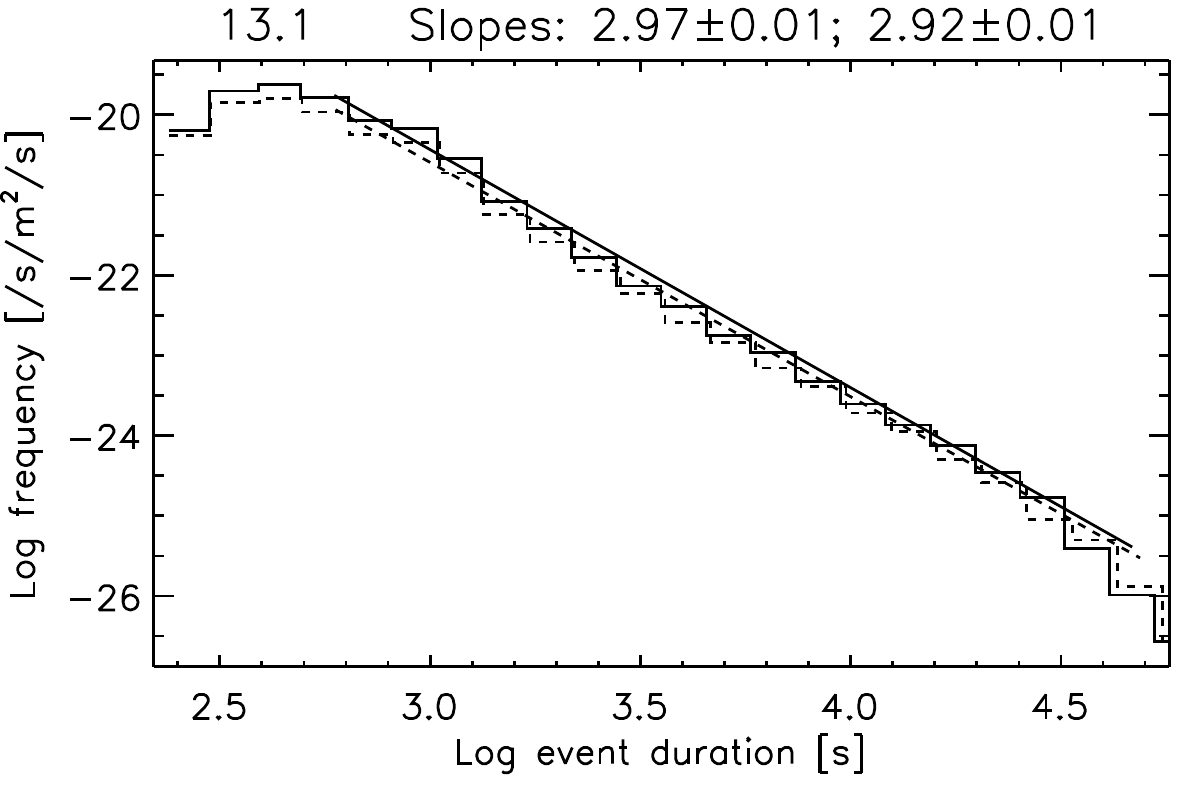}
  \includegraphics[width=.32\linewidth]{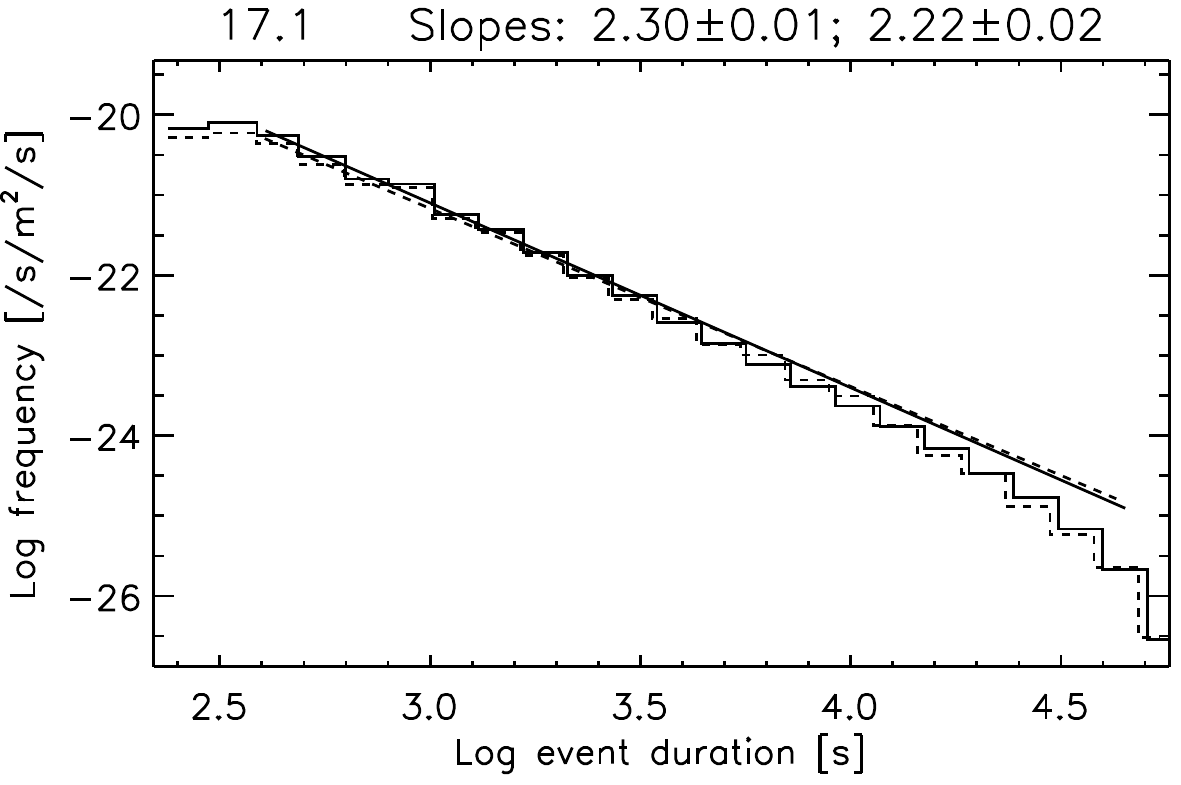}
  \includegraphics[width=.32\linewidth]{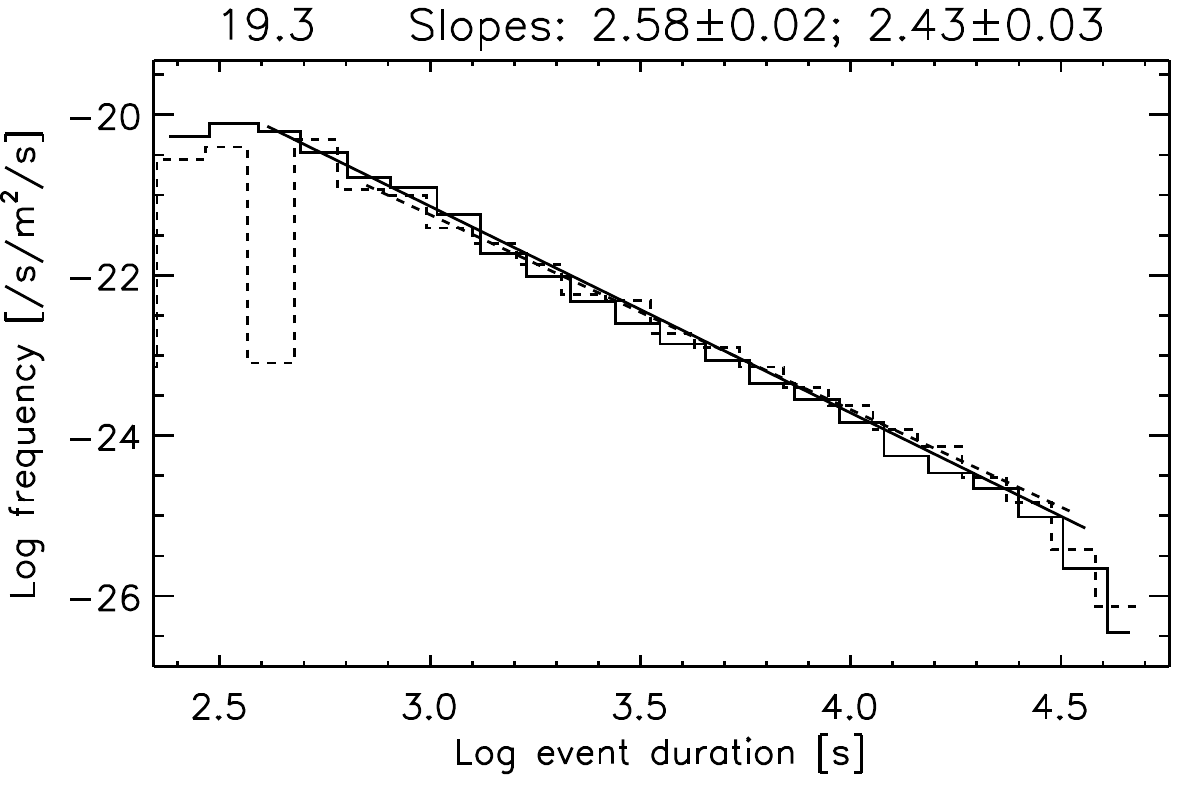} \\
  \includegraphics[width=.32\linewidth]{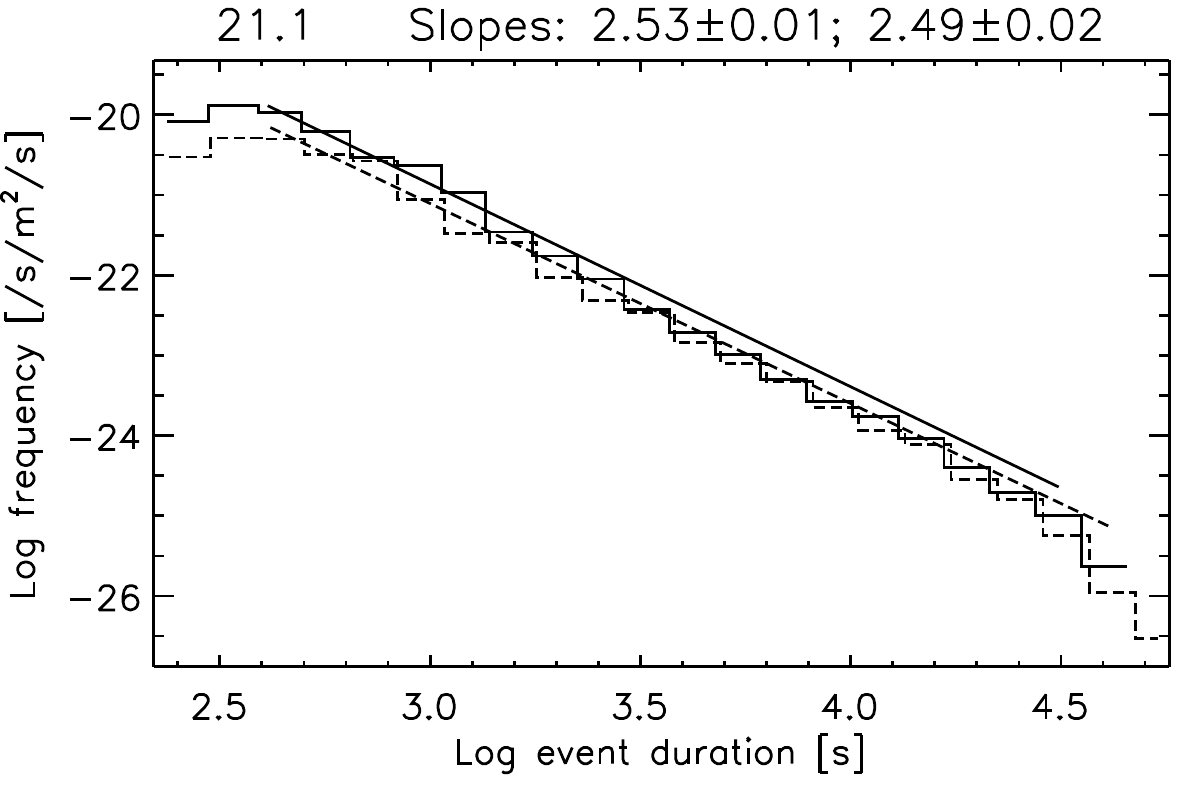}
  \includegraphics[width=.32\linewidth]{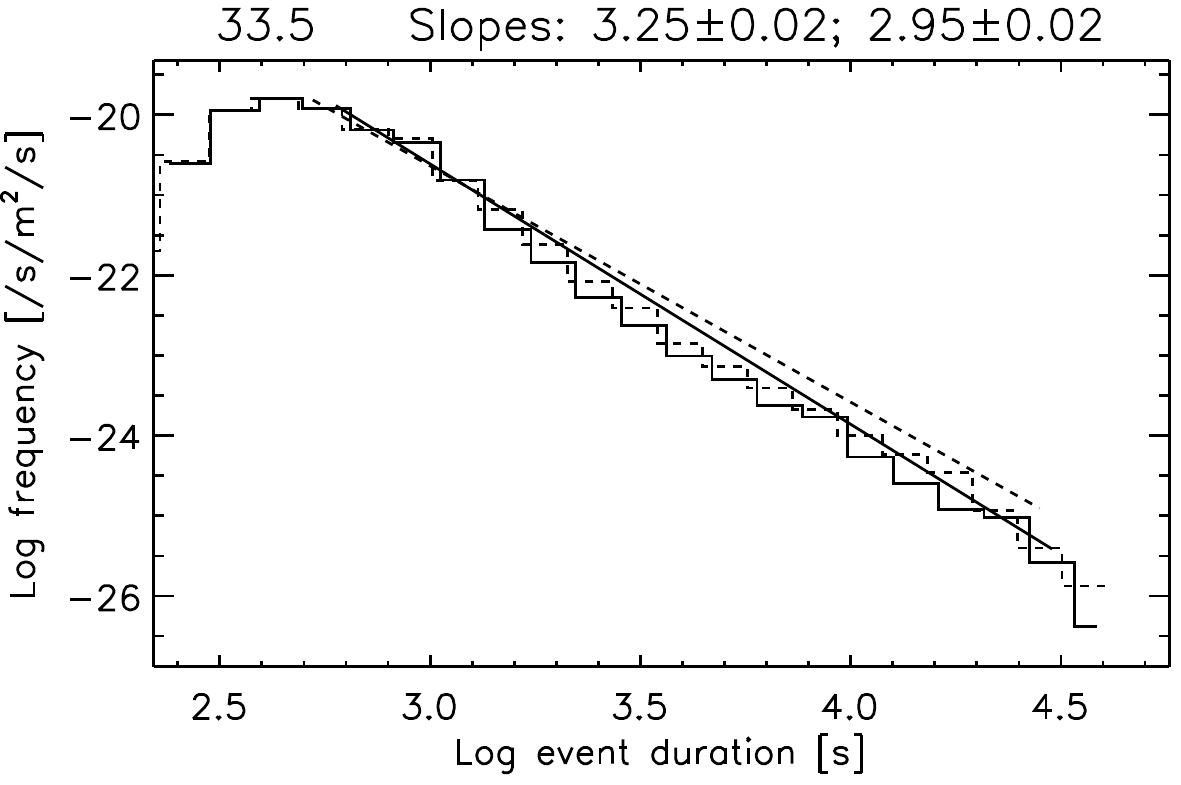}
  \includegraphics[width=.32\linewidth]{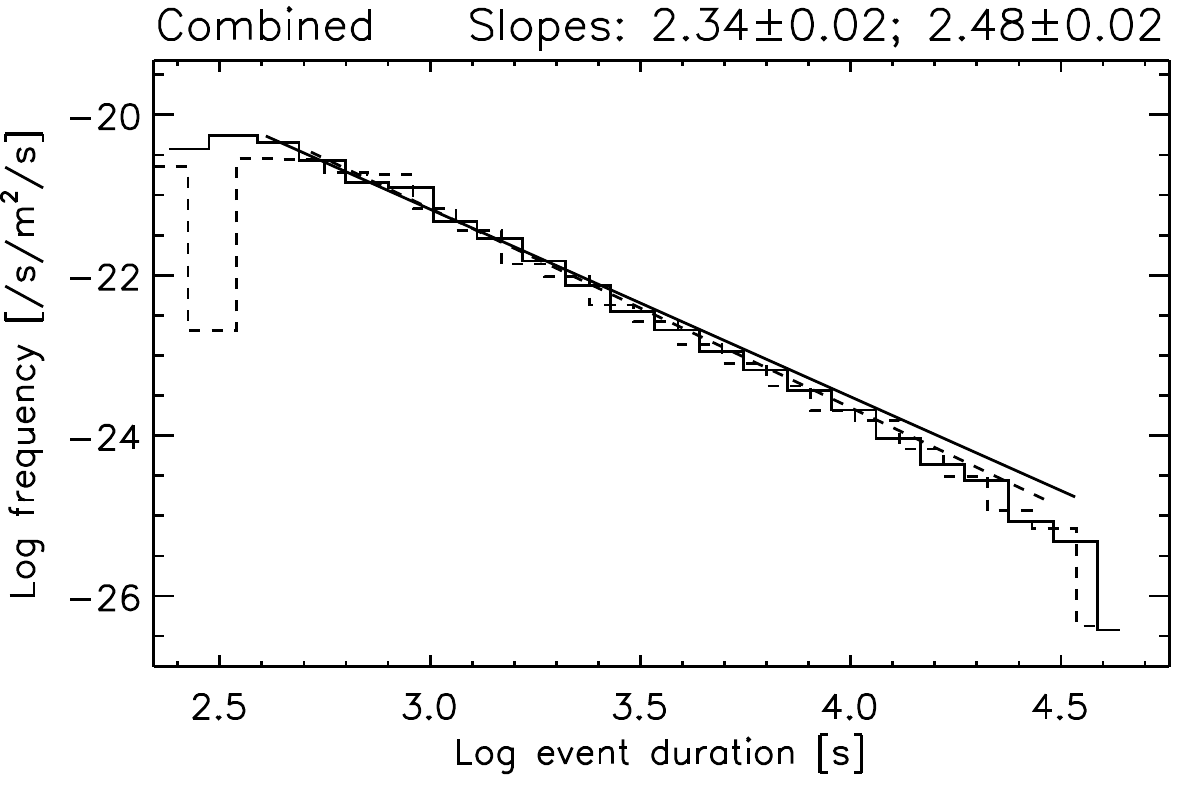}
  \caption{Same as Fig.~\ref{fig:distareas} for the event durations $D$.}
  \label{fig:distdurations}
\end{figure*}

The frequency distributions of event durations $D$ (Fig.~\ref{fig:distdurations}) are also power-laws, extending from the detection threshold of $4\unit{min}$ (2 images) to about $10$ to $17\unit{hr}$, depending on the type of events and observing period.
One could then expect that there is a bias towards a lower number of detections for the longest events, that is the ones of which the duration is significant compared to the observing period of $24\unit{hr}$.
The frequency distributions show indeed that these detected events are less frequent than what would be expected from extrapolations of the power-law fits to the distributions, but this effect is confined to the last points in the distributions, which have a small weight in the fit anyway.

The power-law slopes are between $2.22$ and $2.58$ for events detected in $17.1$, $19.3$, and $21.1\unit{nm}$ and for combined events, and steeper (between $2.92$ and $3.25$) for events detected in the $13.1$ and $33.5\unit{nm}$ bands, which are hotter.
Except for events detected in $33.5\unit{nm}$, the difference between slopes in period~1 and 2 remains small.

\paragraph{Intensities.}

\begin{figure*}
  \includegraphics[width=.32\linewidth]{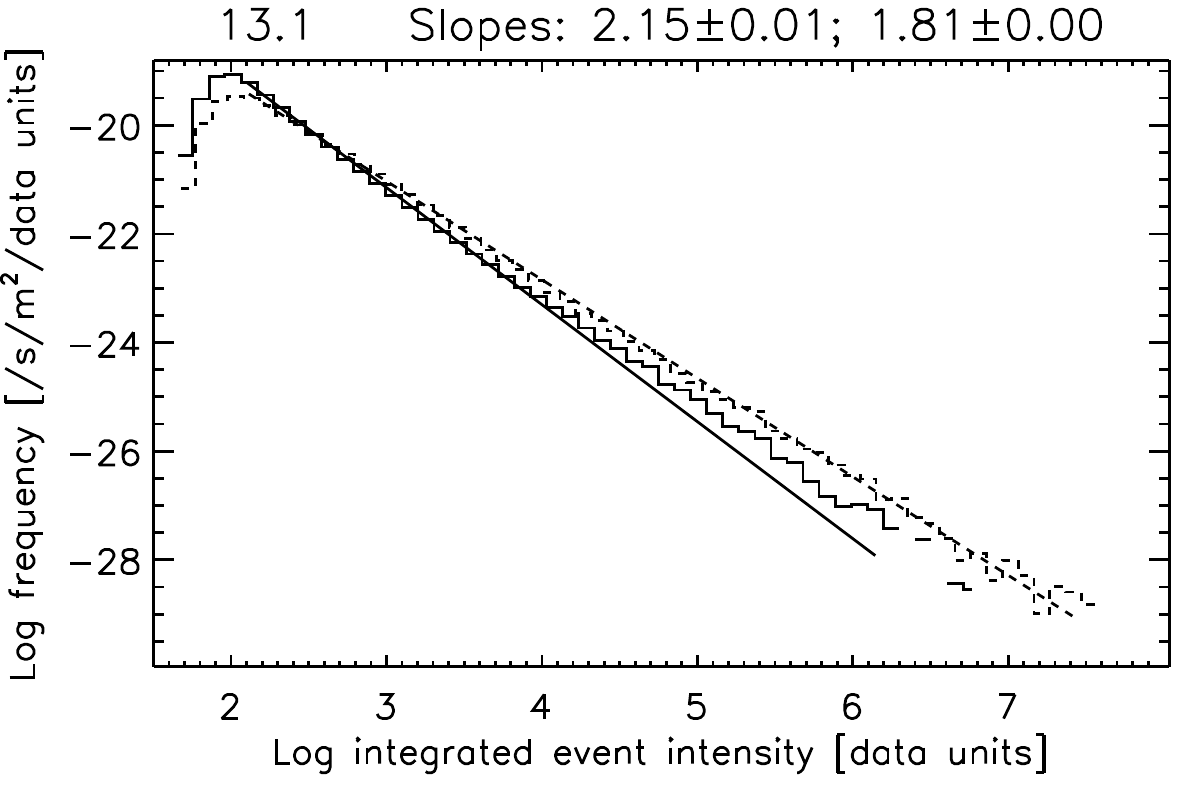}
  \includegraphics[width=.32\linewidth]{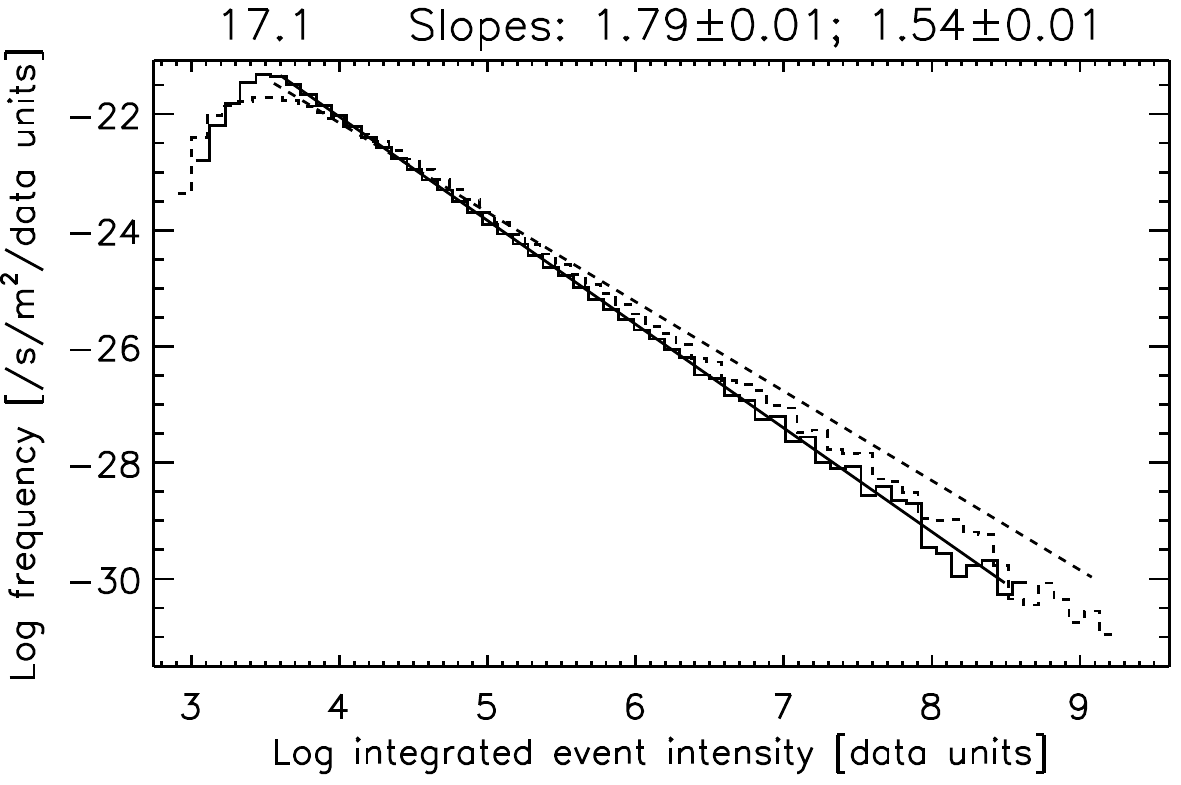}
  \includegraphics[width=.32\linewidth]{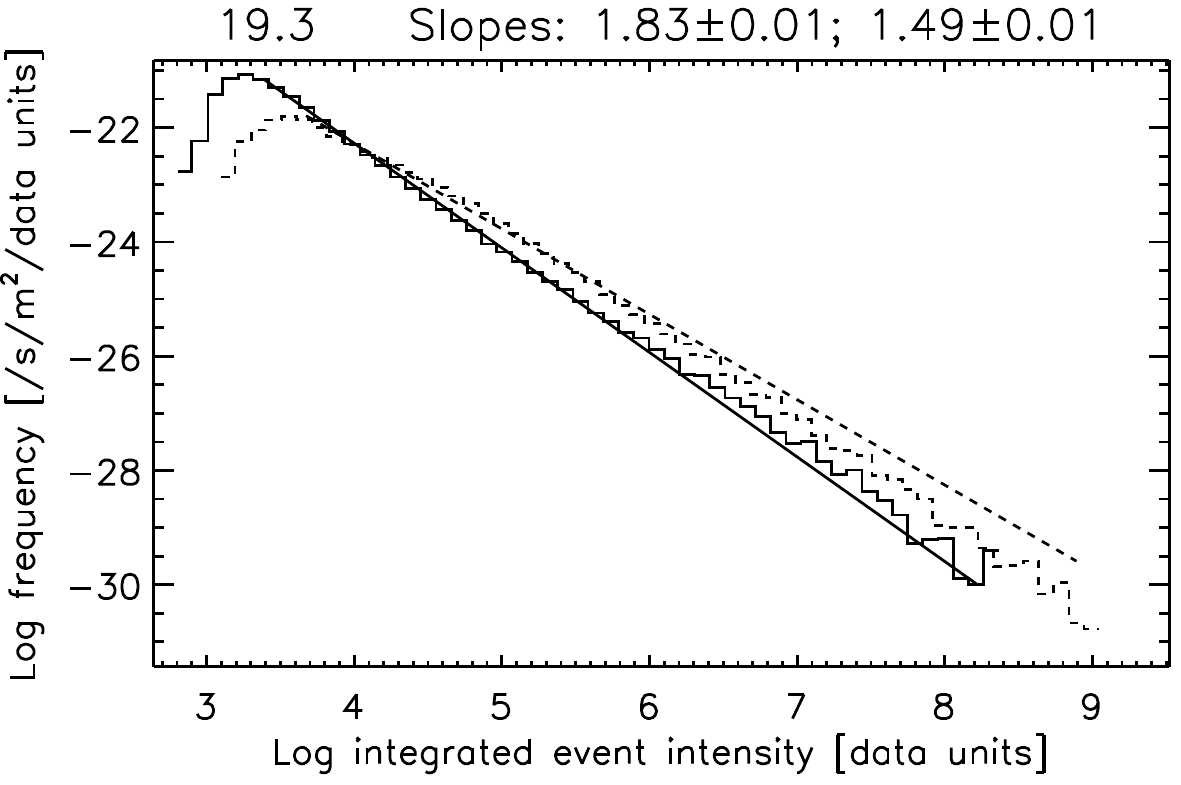} \\
  \includegraphics[width=.32\linewidth]{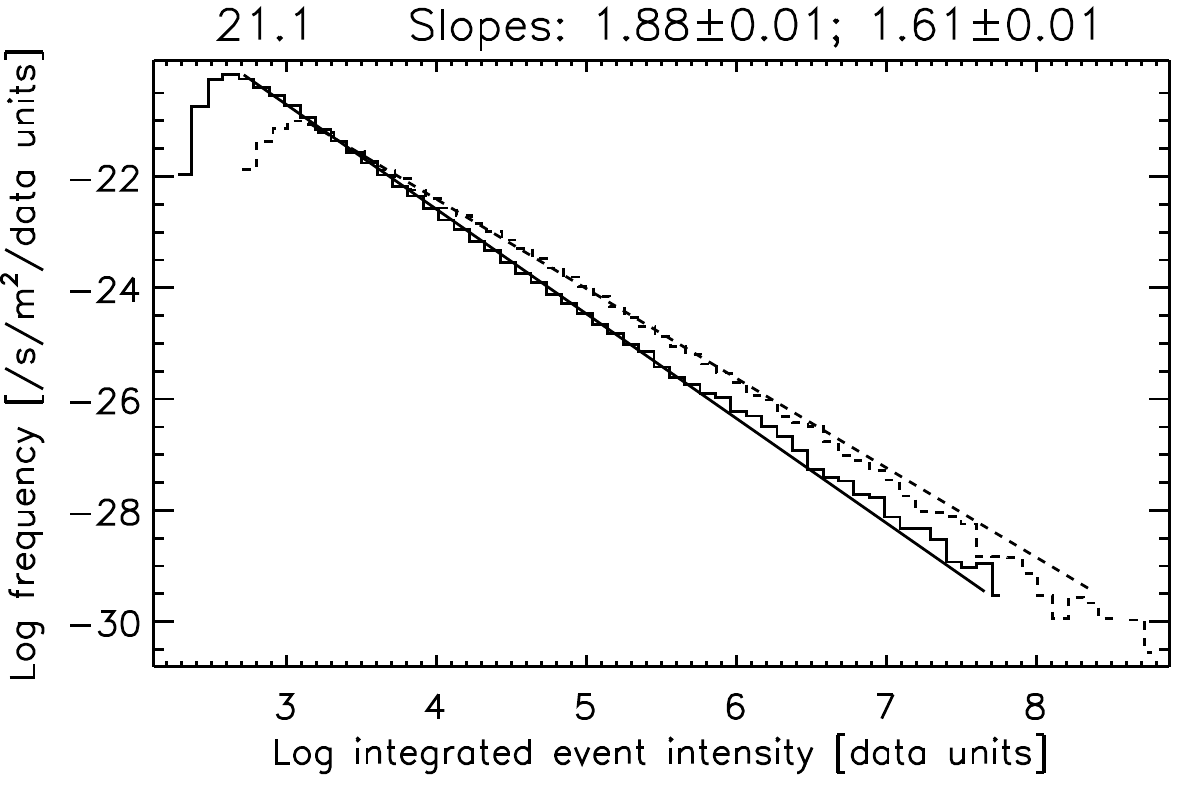}
  \includegraphics[width=.32\linewidth]{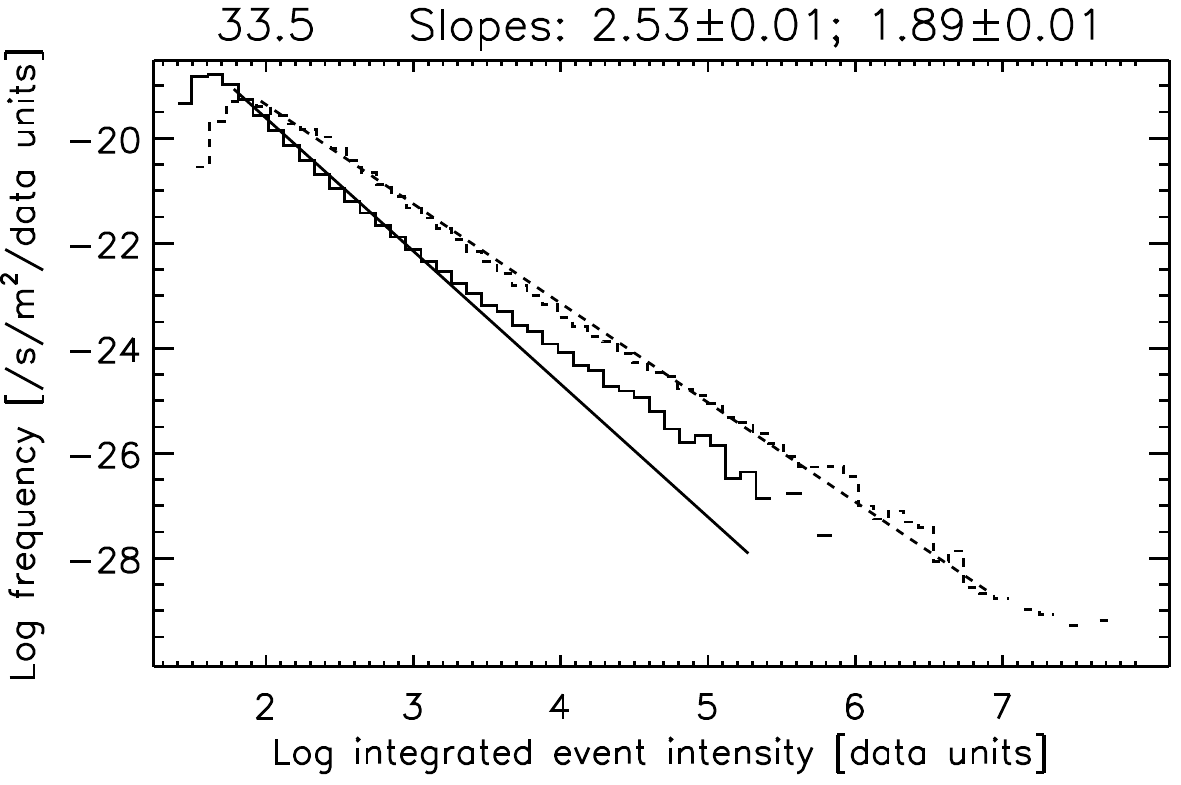}
  \caption{Same as Fig.~\ref{fig:distareas} for the summed intensities for events in each band (this quantity is not relevant for the combined events).}
  \label{fig:distint}
\end{figure*}

The frequency distributions for event total intensities \Itot, shown in Fig.~\ref{fig:distint}, are power-laws over 5 to 6 orders of magnitude in all bands (this quantity is not relevant for the combined events).
The smaller event intensities in $13.1$ and $33.5\unit{nm}$ reflect the lower intensities per pixel in these bands (see Table~\ref{tab:numbers}); the slopes in these bands are steeper than in other bands.
The range in these hotter bands is also reduced to about 4 orders of magnitude in period~1 (Quiet Sun), as the maximum event intensity is smaller than in period~2.
Small intensity events are as frequent in period~1 as in period~2, but higher intensity events are less frequent in period~1 than in period~2; as a result, the fitted slope (in which the large number of small intensity events has more weight) is steeper in period~1 (between $1.79$ and $1.88$ in period~1, compared to $1.49$ to $1.61$ in period~2 in the $17.1$, $19.3$, and $21.1\unit{nm}$ bands).

\paragraph{Maximum intensities.}

\begin{figure*}
  \includegraphics[width=.32\linewidth]{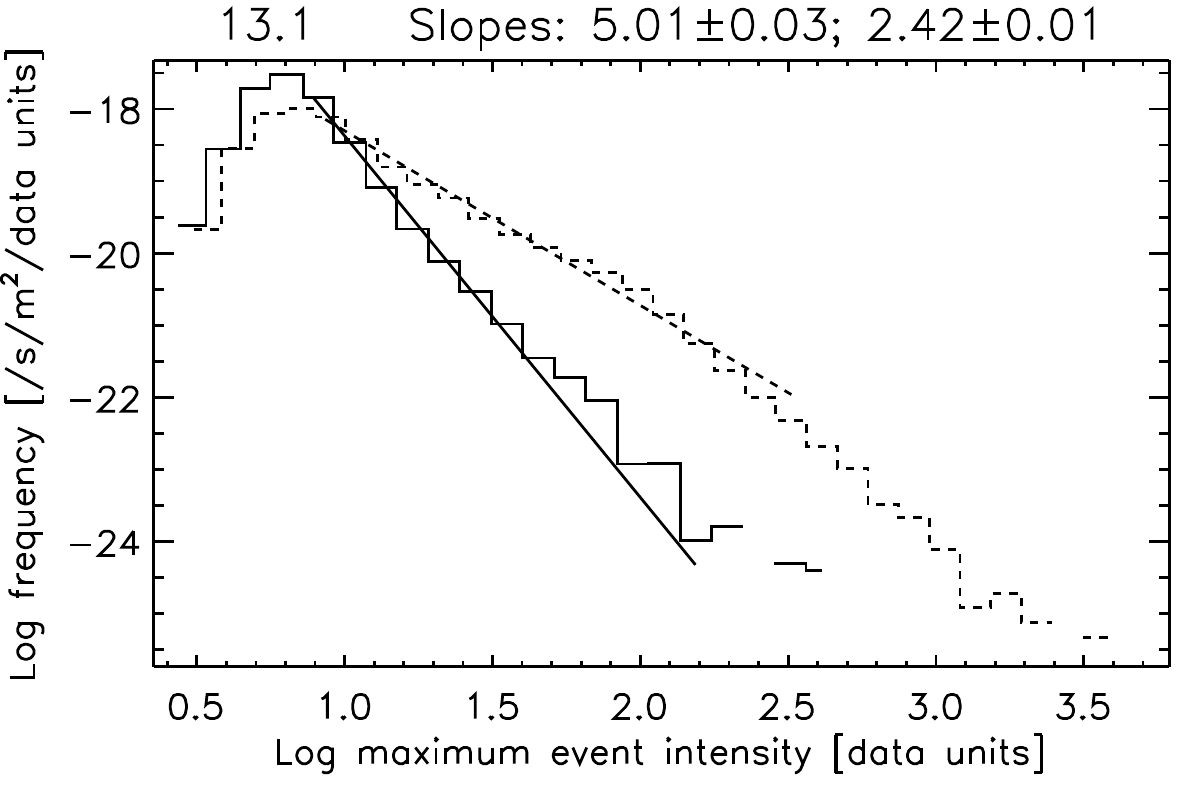}
  \includegraphics[width=.32\linewidth]{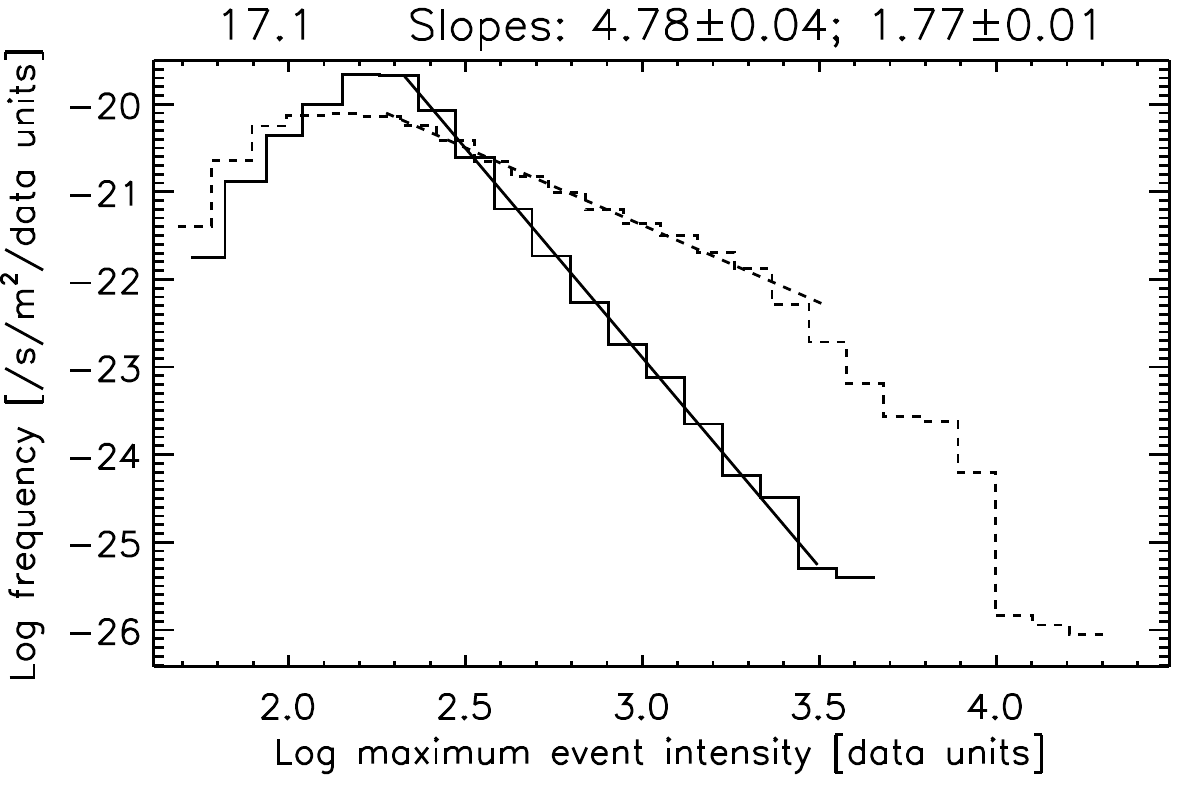}
  \includegraphics[width=.32\linewidth]{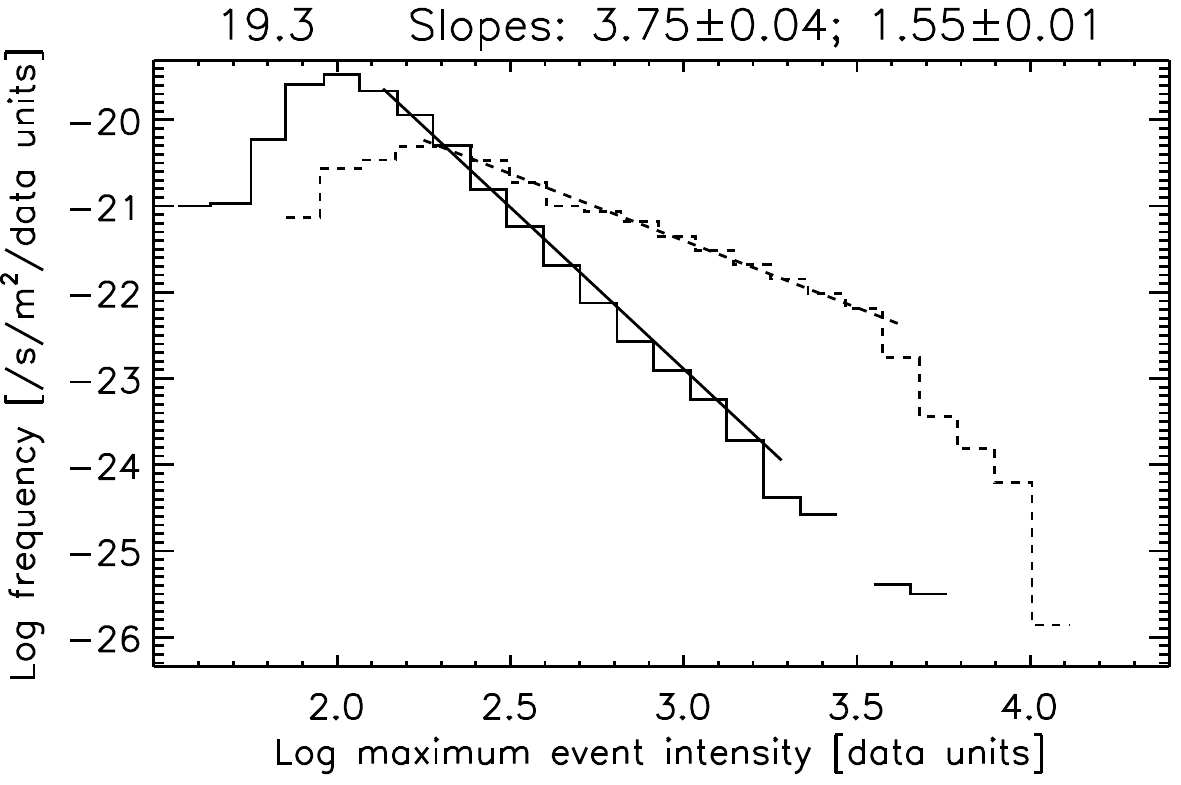} \\
  \includegraphics[width=.32\linewidth]{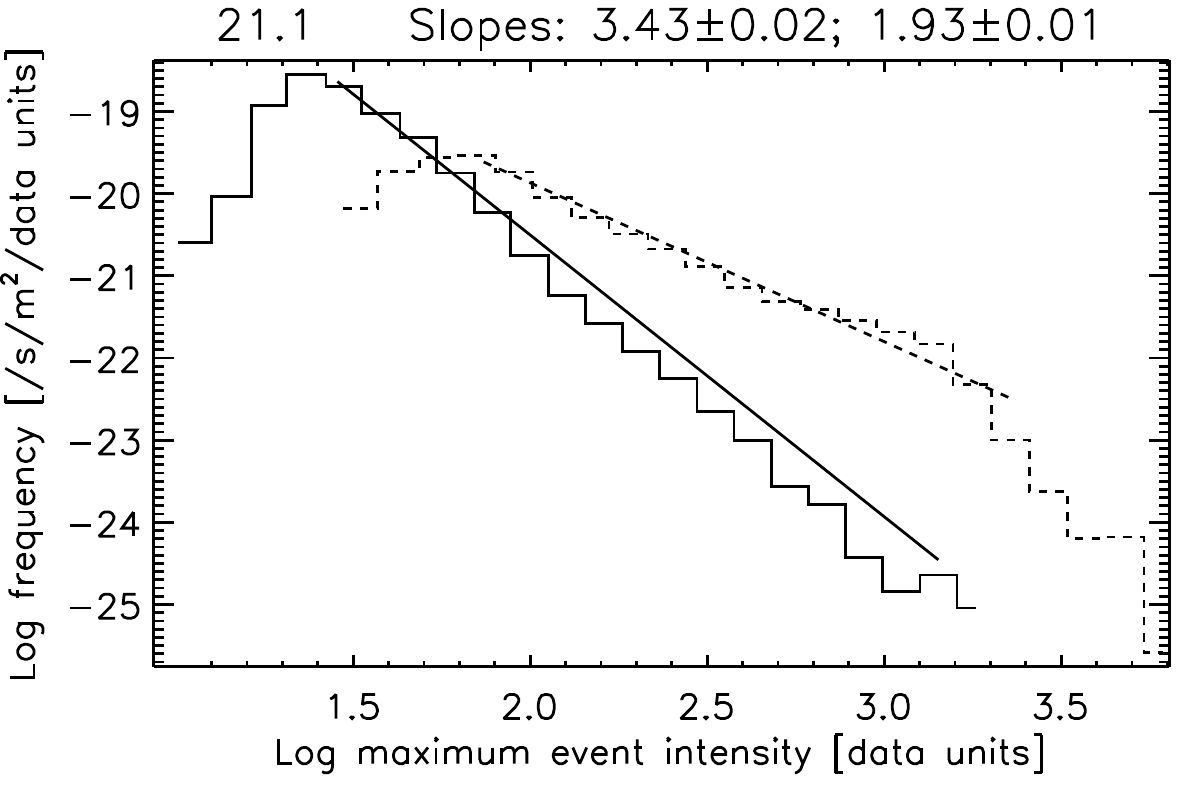}
  \includegraphics[width=.32\linewidth]{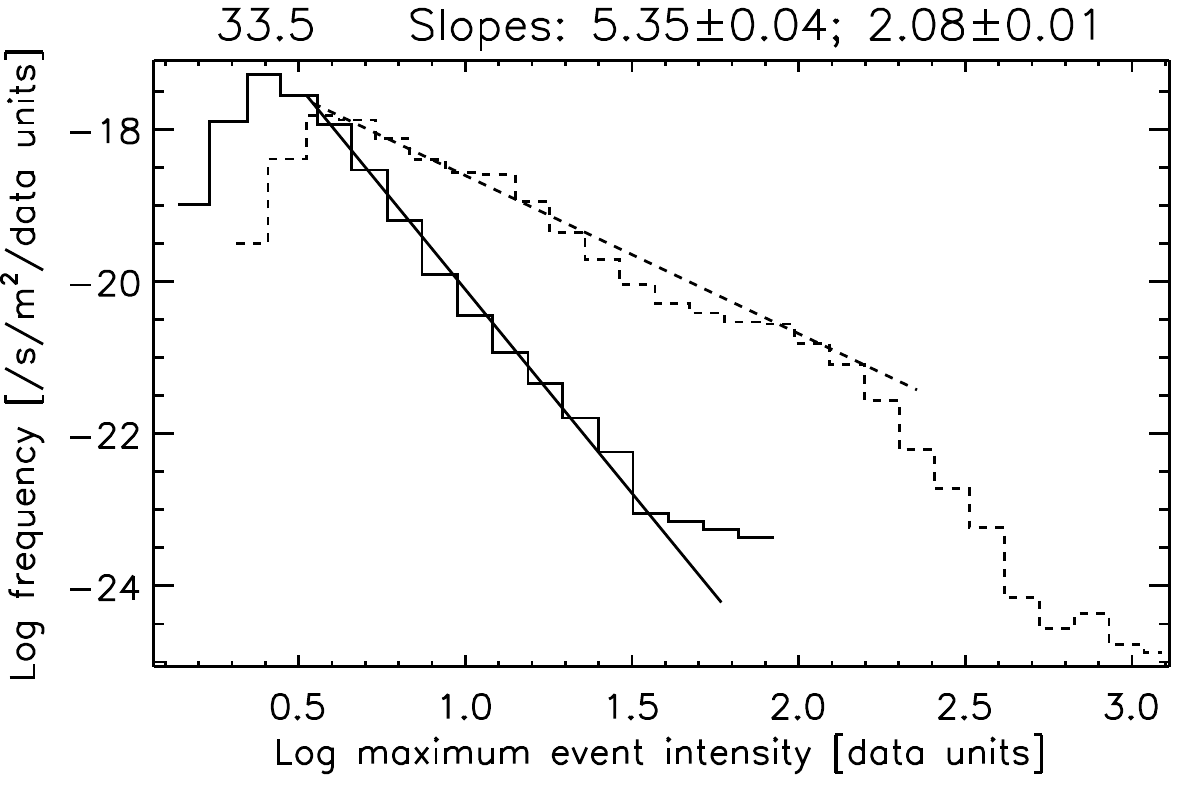}
  \caption{Same as Fig.~\ref{fig:distareas} for the maximum intensities for events in each band (this quantity is not relevant for the combined events).}
  \label{fig:distintmax}
\end{figure*}

The value ranges are reduced and the slopes of the distributions are steeper for maximum intensity \Imax (Fig.~\ref{fig:distintmax}) than for total intensity \Itot, especially in period~1, when slopes reach a value of more than 5 in some cases.
The reduced ranges and higher slopes compared to total intensity can be explained by the influence of the dispersion of event durations on the distribution of total intensity (a sum over event duration) compared to maximum intensity, which is the intensity at one given time only.

\paragraph{Thermal energies.}

\begin{figure*}
  \includegraphics[width=\linewidth]{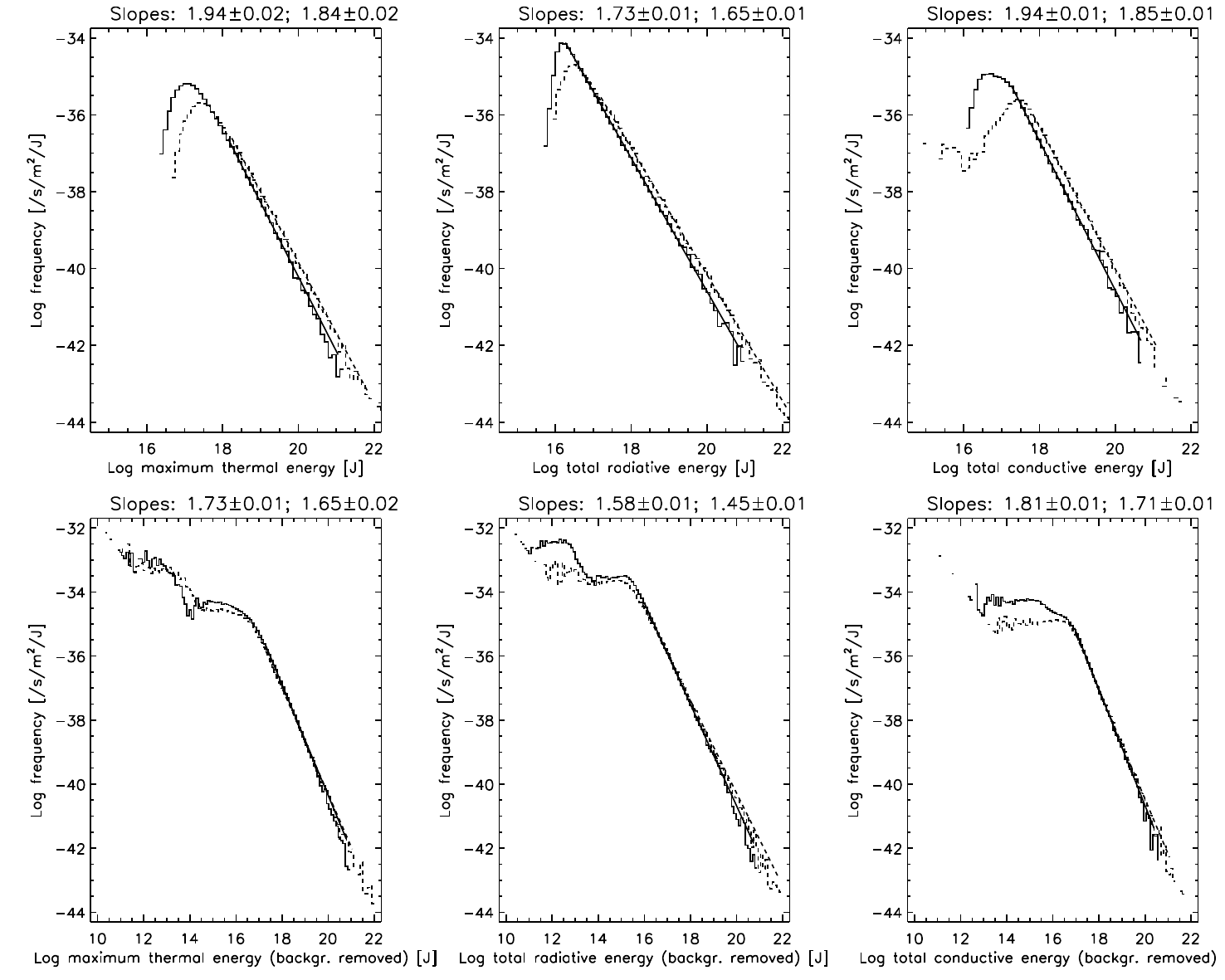}
  \caption{Frequency distributions of thermal, radiative, and conduction energies (from left to right) for the combined events, for period~1 (plain lines) and period~2 (dashed lines). The top panels are for quantities before background subtraction and the bottom panels are for background-subtracted quantities. }
  \label{fig:disten}
\end{figure*}

The frequency distributions of maximum thermal energies \Eth\ (without subtraction of background) and \Ethb\ (with subtraction of background) are represented in Fig.~\ref{fig:disten}.
The highest energy is about $10^{21}\unit{J}$ for period~1 and $10^{22}\unit{J}$ for period~2: the highest energies correspond to large microflares and are only moderately affected by background subtraction.
In the low-energy range however (as low as a few $10^{16}\unit{J}$ for \Eth, corresponding to small nanoflares), background subtraction can reduce the event energy by a large factor.
This widens and flattens the \Ethb\ distribution at small energies, as can be seen in the bottom panels of Fig.~\ref{fig:disten}, and the same effect can lead to negative values of the background-subtracted energy in some cases (not represented).
This is because events are (and can only be) defined from variations in intensity.
However, this affects only the smallest-energy events, and the difference between slopes in the cases with and without background subtraction gives an upper bound for this effect in the power-law ranges.

Even for the thermal energies with no background subtraction \Eth, the number of events in the very low-energy range is reduced as a result of the thresholds in event duration and volume: because of the imperfect correlation between these parameters used for thresholds and the thermal energy (correlations are shown in detail in Sec.~\ref{sec:correlations}), some real events that are not considered here (because they are below the duration and volume thresholds) could have an energy higher than $10^{16}\unit{J}$ but are missing in our statistics.
As a result, the very low-energy range of distribution of \Eth\ is biased towards smaller frequencies, and this explains that the power-law does not extend as far as the smallest energies of detected events.

For these reasons, we start the fitting ranges at $2\times10^{18}\unit{J}$ for \Eth\ and $2\times10^{17}\unit{J}$ for \Ethb, so that these biases do no affect the power-law fits.
We obtain slopes of $1.94$ and $1.84$ for \Eth\ in both observing periods, and $1.73$ and $1.65$ for \Ethb: the slopes are all less than 2. The slopes are lower for period~2 (when there are comparatively more high-energy events) and when the background is subtracted.

\paragraph{Radiative energies.}

Radiative energies of events cover about the same range as thermal energies, with a power law distribution from $10^{16}$ to $10^{21}\unit{J}$ in period~1 and to $10^{22}\unit{J}$ in period~2.
For \Eradb\ (with background subtraction), we notice the same biases at low energy than for $\Ethb$.
On the contrary, the distribution of \Erad\ remains a power law almost until the smallest energies.
We obtain slopes of $1.73$ and $1.65$ for \Erad\ in both observing periods, and $1.58$ and $1.45$ for \Eradb.
These slopes are significantly less than 2, and they have the same behaviour as the slopes for \Eth\ and \Ethb: they are lower for period~2 and when the background is subtracted.

\paragraph{Conduction energies.}

The distributions of conduction energies also cover about the same range as thermal and radiative energies, also the power-law regime range is reduced by about one order of magnitude for small energies.
Like for other kinds of energies, background subtraction makes the distribution of \Econdb\ flat at low energies.
The slopes are $1.94$ and $1.85$ for \Econd\ in both observing periods, and $1.81$ and $1.71$ for \Econdb.

\subsection{Correlations between event parameters}
\label{sec:correlations}

\begin{figure}
  \includegraphics[width=\linewidth]{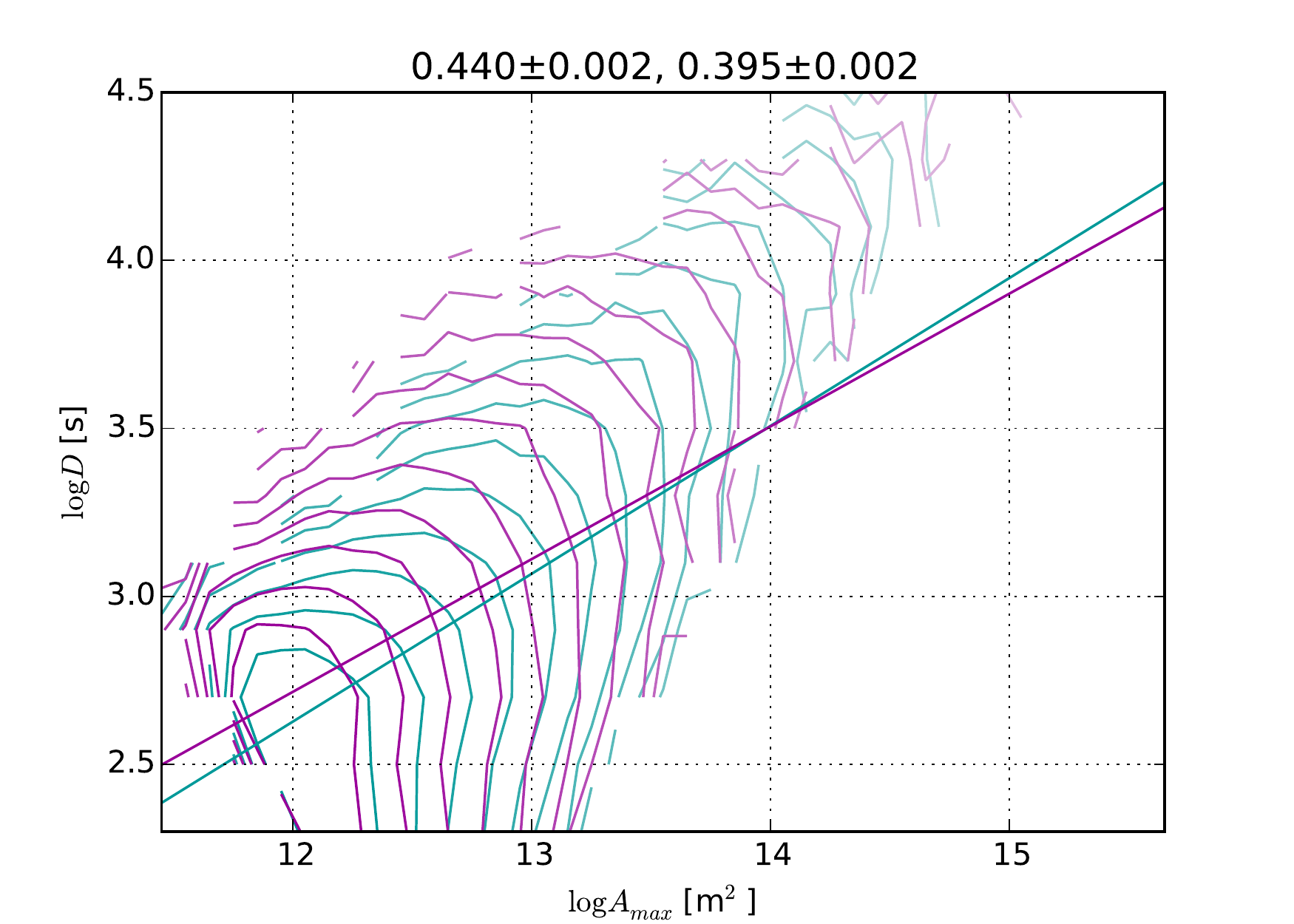}
  \caption{Joint distribution of the combined events maximum area \Amax\ and their duration $D$, for periods 1 (cyan) and 2 (magenta). The curves are level lines of the joint empirical probability distribution of \Amax\ and $D$, obtained by dividing a 2D histogram of these quantities by the number of data points and by the area of each bin (bins are spaced exponentially); probabilities associated to different curves are spaced exponentially and are separated by a half decade, from low probabilities (light colours) to high probabilities (dark colours). Linear fits of $D(\Amax)$ are overplotted, and their slopes are shown above the plot.}
  \label{fig:cad}
\end{figure}

\paragraph{Areas and durations.}
The correlations between maximum event area \Amax\ and duration $D$ (for combined events) are shown in Fig.~\ref{fig:cad}, as represented by the joint frequency distributions of $\Amax$ and $D$, in both observing periods.
We model each correlation by a power law, that we determine from a linear fit of all the $(\log\Amax,\log D)$ pairs.
By considering that all points have a $0.2$-decade uncertainty, we obtain an uncertainty on the slope of order $2\;10^{-3}$, a low value due to the large number of points used for each fit (see Table~\ref{tab:numbers}).
However, the power-law model does not represent the data perfectly, and these uncertainties are then not entirely relevant.

We obtain a slope of $0.39$ for period~1 and $0.44$ for period~2.
However, the correlation is dominated by the large number of small events, and the slopes would be about $0.8$ and $0.7$ if only large events ($\Amax>10^{13}\unit{m^2}$) were considered.
One tentative interpretation for this slope increase at larger areas is that the duration is actually proportional to the loop half-length $L$ (as justified by time scale computations as in \citet{auchere14a}), with a loop aspect ratio $a(L)$ increasing with loop length: then the area is proportional to $L^2/a(L)$, yielding a steeper slope of the area-duration correlation for larger areas.
There is also more dispersion for period~2.

\paragraph{Areas and energies.}

\begin{figure*}
  \includegraphics[width=\linewidth]{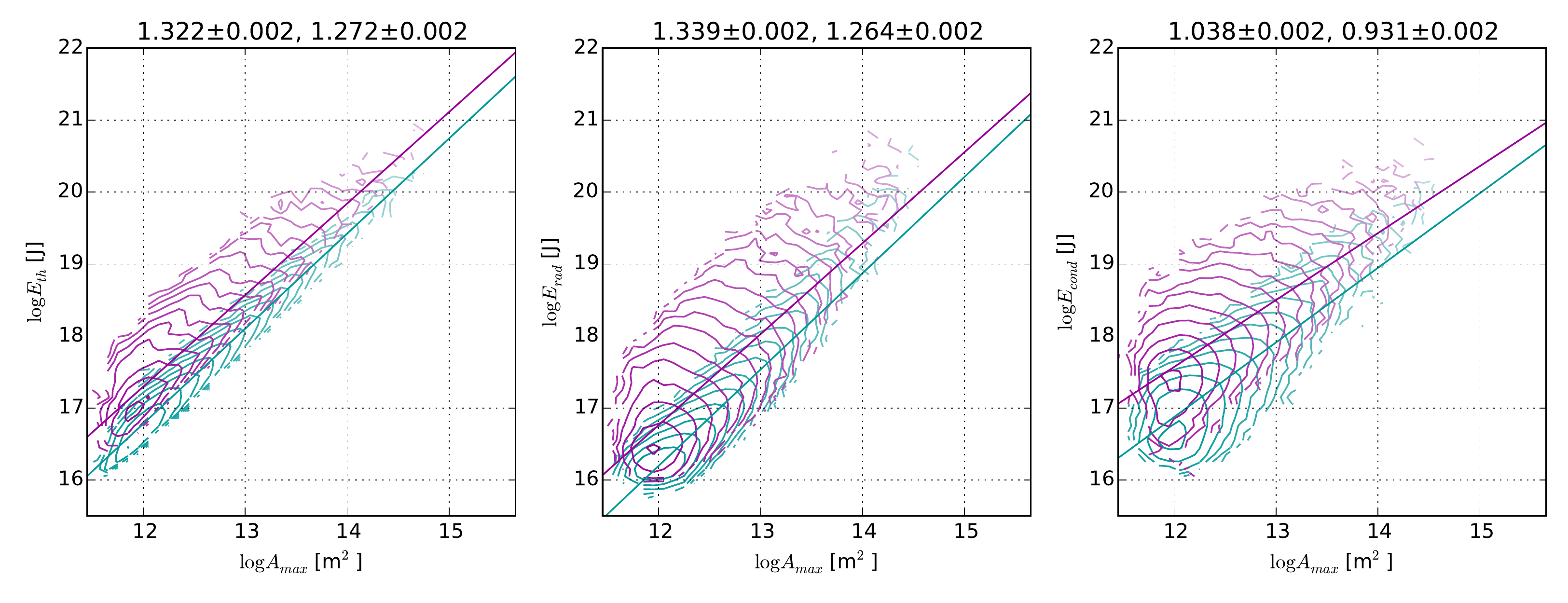}
  \caption{Joint distribution of the maximum area \Amax\ and the energies (from left to right: maximum thermal energy, radiative energy, and conduction energy) of combined events, without background subtraction, with the same plotting conventions as in Fig.~\ref{fig:cad}.}
  \label{fig:cae}
\end{figure*}

The correlations between maximum event area \Amax\ and energies (for combined events) are shown in Fig.~\ref{fig:cae}.
The slopes of power-law fits to these correlations are 1.32 (for thermal energy) and 1.34 (for radiated energy) in period~1 (quiet Sun), and have slightly smaller values (1.27 and 1.26 respectively) in period~2 (active Sun).
The same difference is seen in the correlations between areas and conduction energies, although slopes are smaller (1.04 in period~1 and 0.93 in period~2).
A more striking feature of these correlations is that they are shifted towards higher energies and broader in period~2: for a given area, the average and the standard deviation of energies (thermal, radiated or conducted) are both higher when the observed area is more active.

\paragraph{Durations and energies.}

\begin{figure*}
  \includegraphics[width=\linewidth]{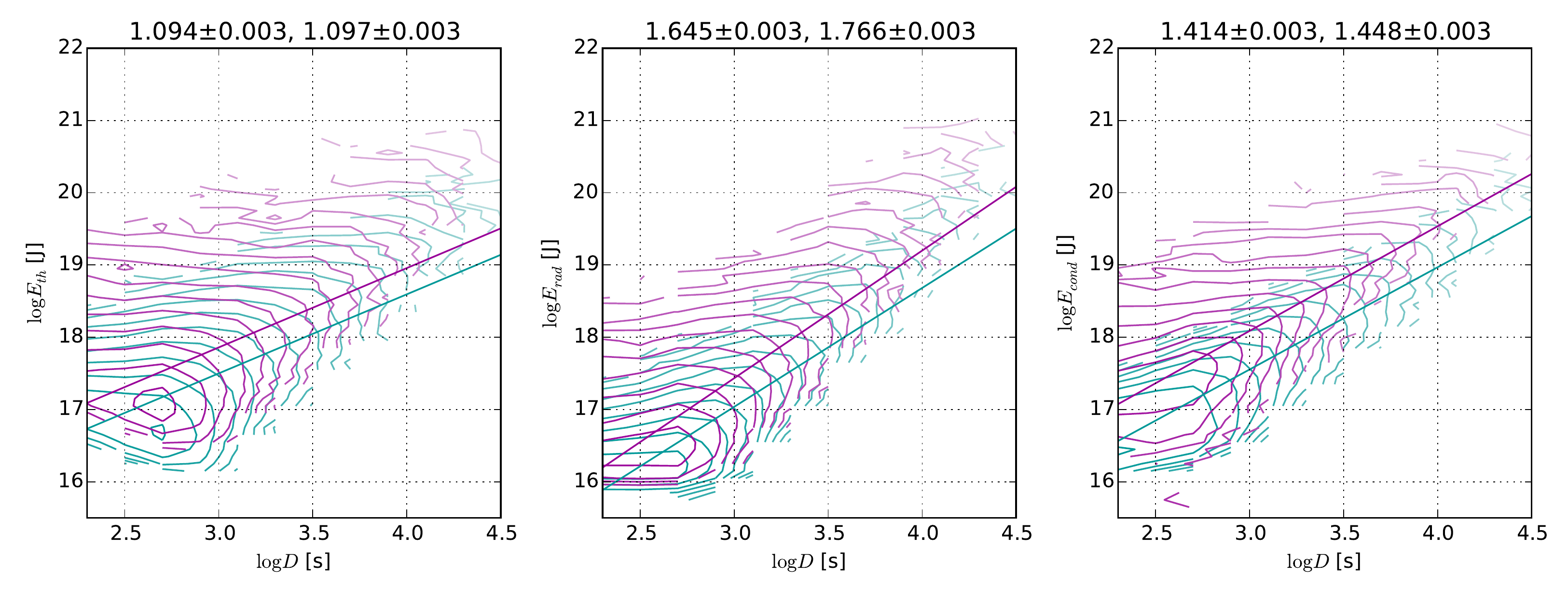}
  \caption{Joint distribution of the duration $D$ and the energies (from left to right: maximum thermal energy, radiative energy, and conduction energy) of combined events, without background subtraction, with the same colour scale as in Fig.~\ref{fig:cad}.}
  \label{fig:cde}
\end{figure*}

The correlations between event duration $D$ and energies (for combined events) are shown in Fig.~\ref{fig:cad}.
Similarly to the areas-energies correlations, the durations-energies correlations are broader and shifted towards higher energies in period~2.
The slopes of power-law fits to these correlations are close to $1.1$ for thermal energies, $1.7$ for radiated energies, and $1.4$ for conducted energies, with slopes somewhat larger in period~2 than in period~1.

\paragraph{Energies.}

\begin{figure*}
  \includegraphics[width=\linewidth]{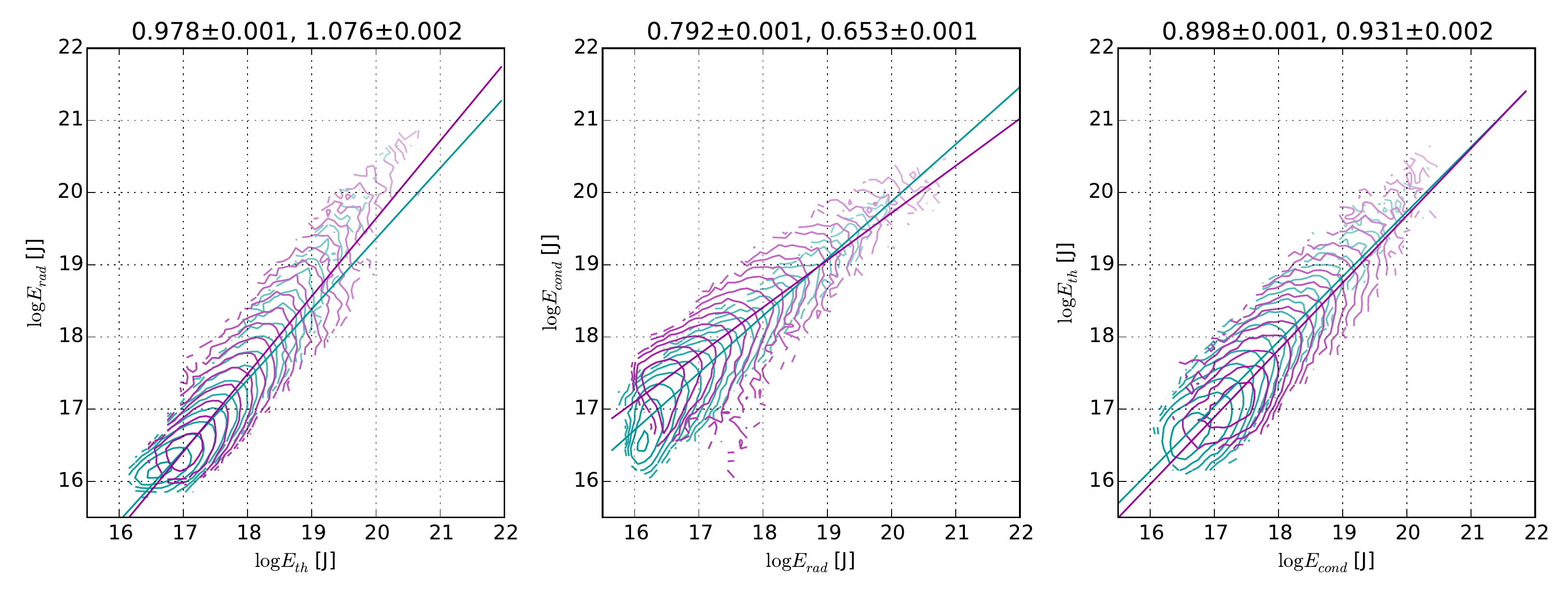}
  \caption{Joint distributions of the combined events energies (without background subtraction), from left to right: maximum thermal energy and radiative energy, radiative energy and conducted energy, and conducted energy and maximum thermal energy, with the same colour scale as in Fig.~\ref{fig:cad}.}
  \label{fig:cee}
\end{figure*}

All event energies (thermal, radiated and conducted) are well correlated (Fig.~\ref{fig:cee}).
Power-law exponents are close to 1 for the $\Eth-\Erad$ correlation, and about $0.9$ for $\Econd-\Eth$, and significantly lower for $\Erad-\Econd$.
The parameters of these power-law fits (exponent and constant factor) confirm that all three energy estimates have the same order of magnitude.
It also becomes apparent in these plots that the minimal thermal and conducted energies in detected events are affected by the observing period (1 or 2), while the minimal radiated energy is not.

\section{Discussion}

\begin{figure*}
  \includegraphics[width=.49\linewidth]{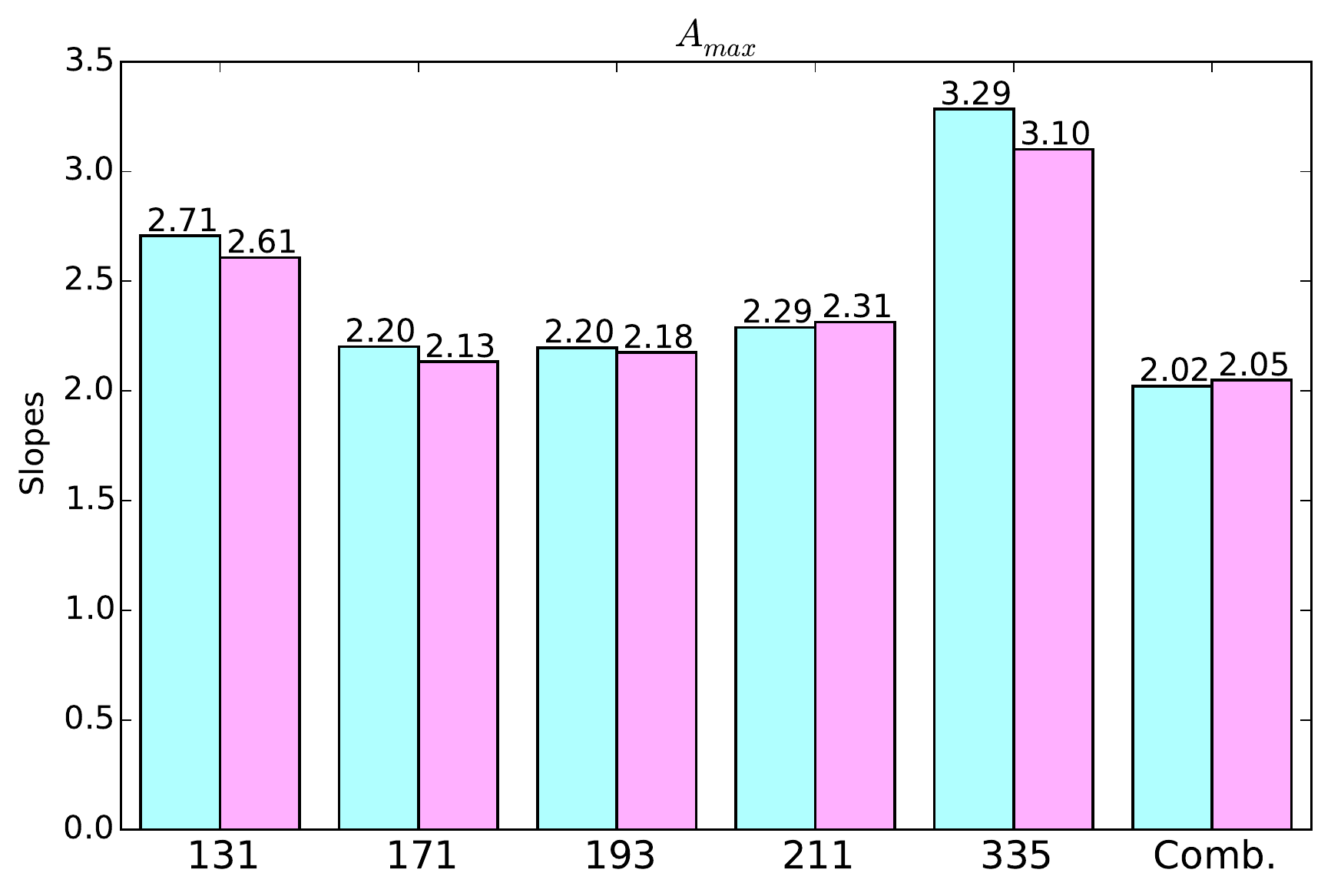}\hfill
  \includegraphics[width=.49\linewidth]{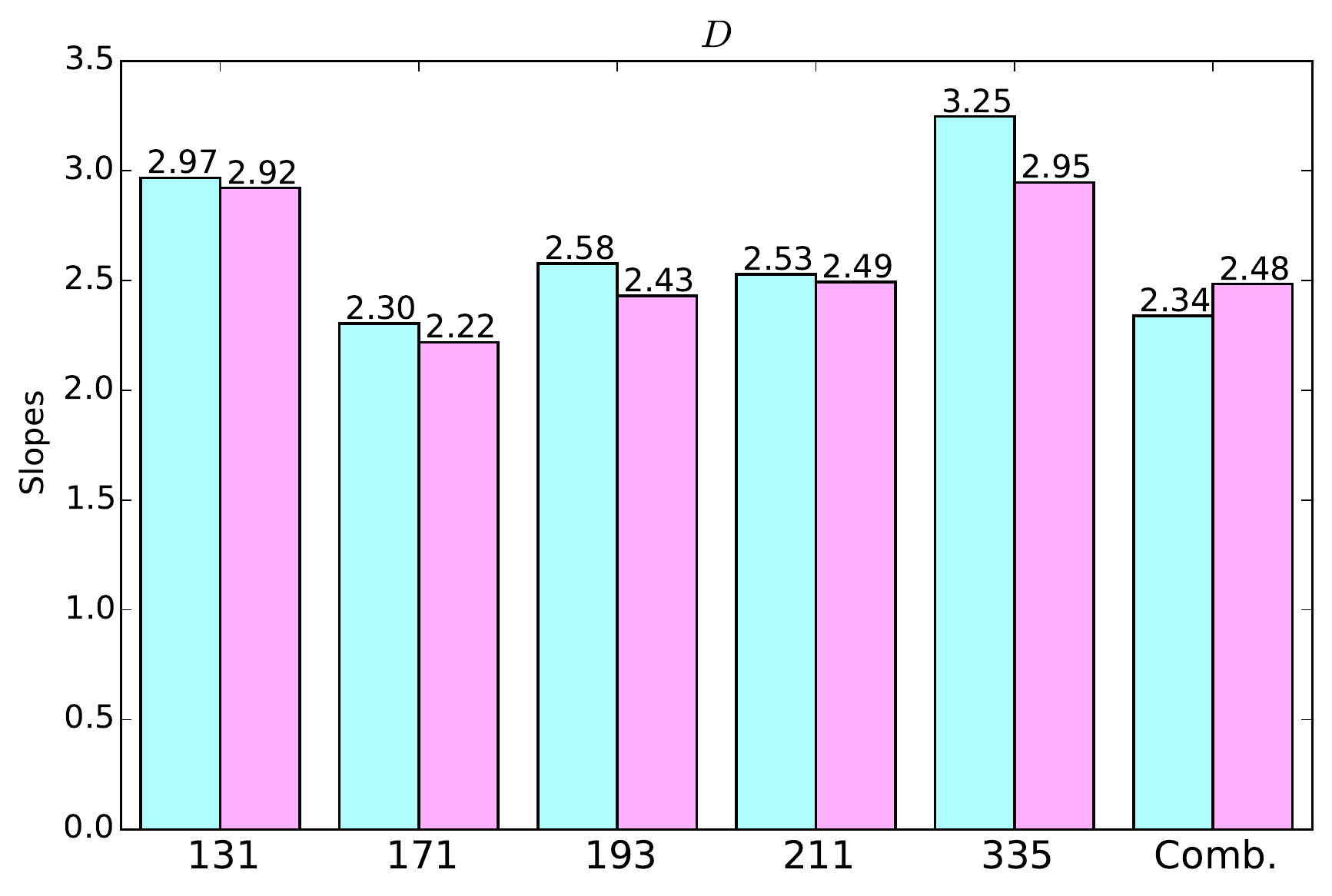}\\
  \includegraphics[width=.49\linewidth]{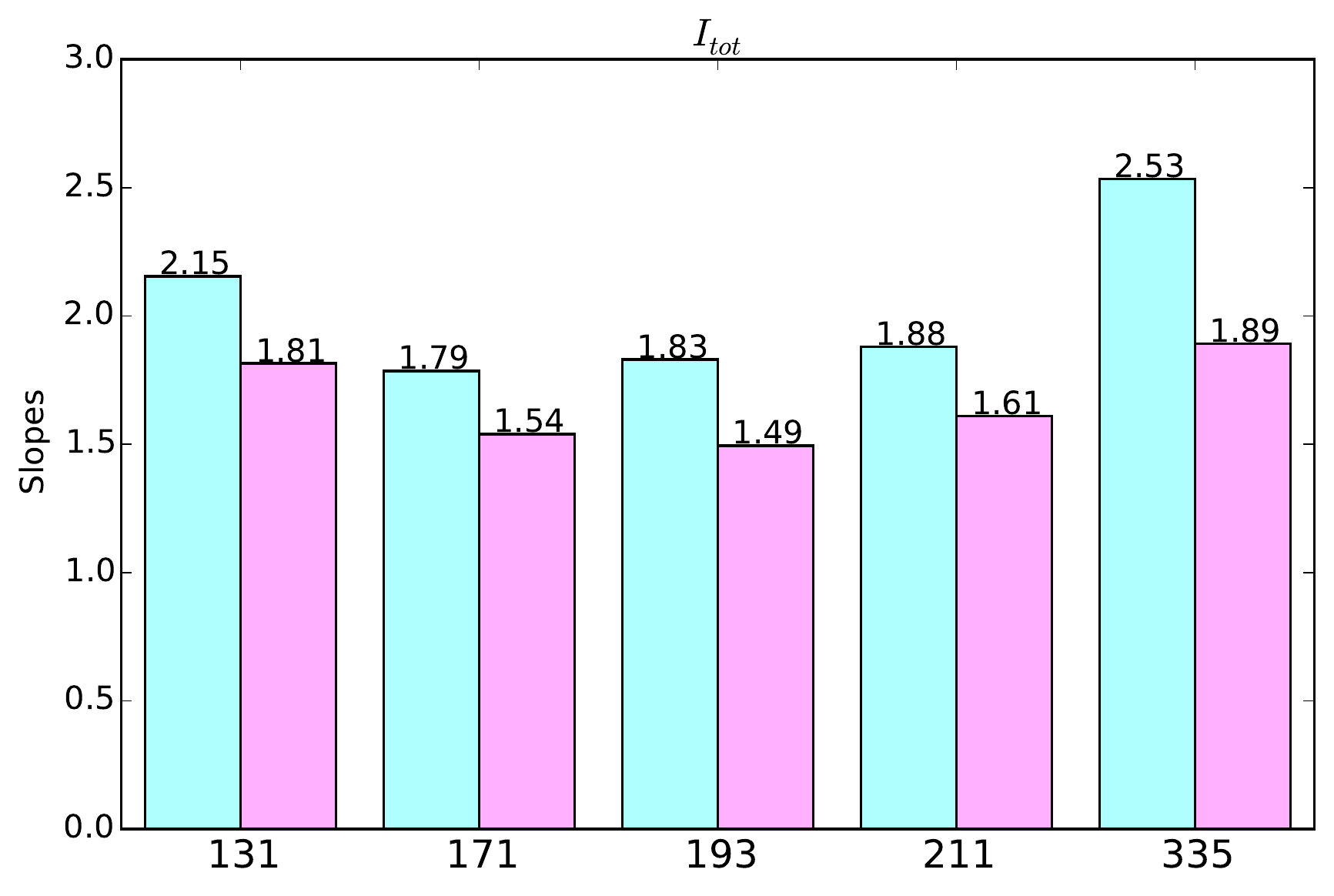}\hfill
  \includegraphics[width=.49\linewidth]{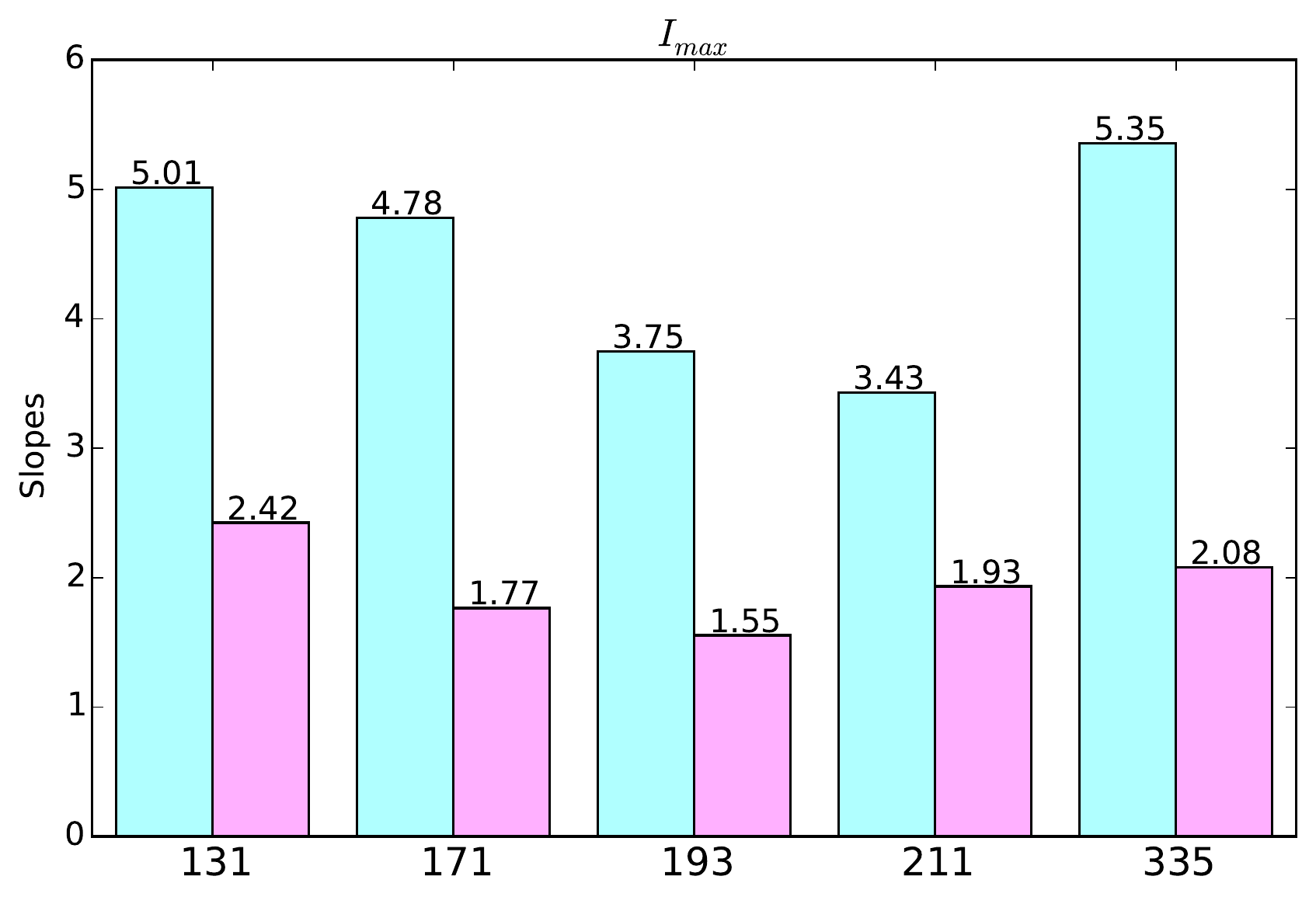}\\
  \includegraphics[width=.49\linewidth]{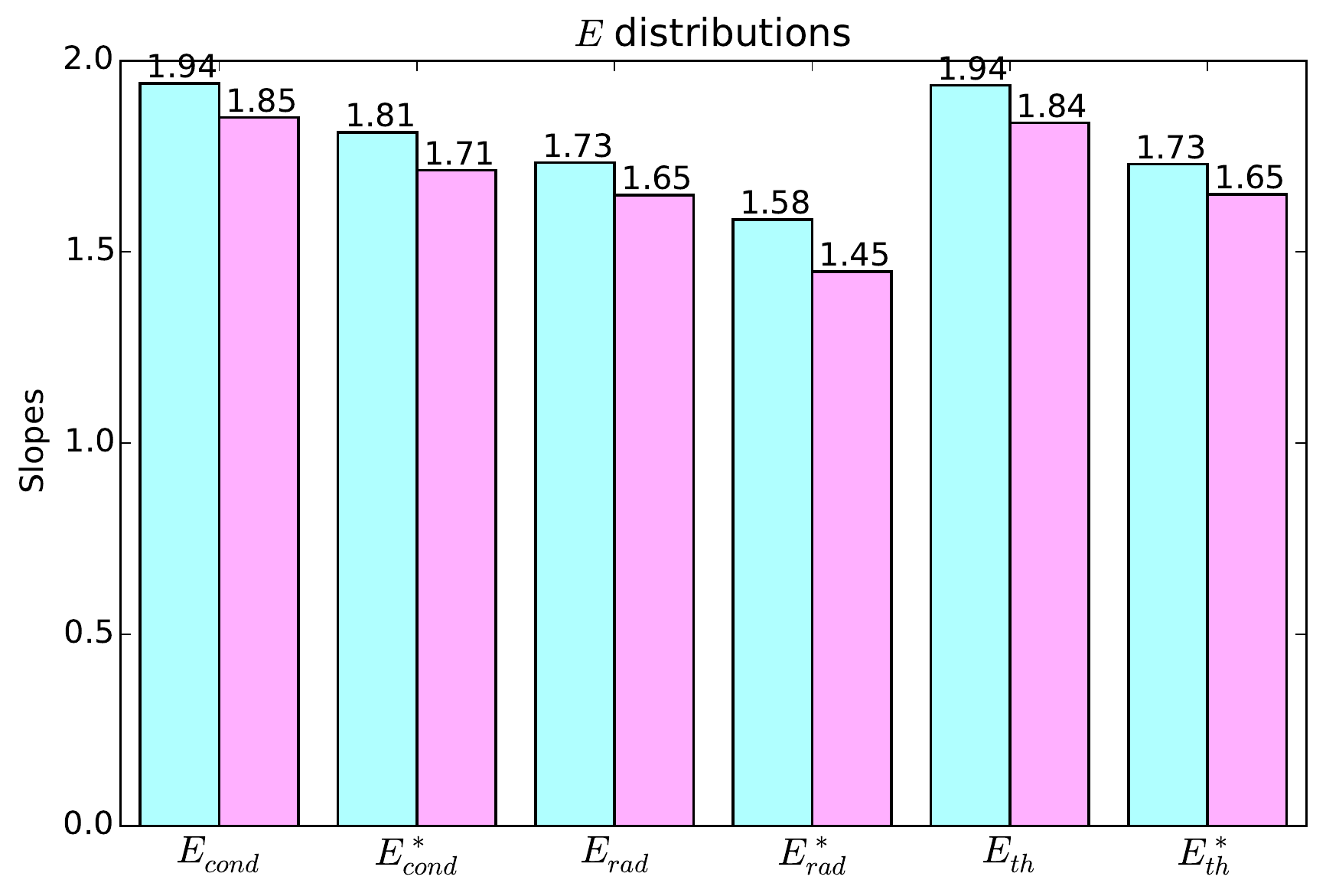}\hfill
  \includegraphics[width=.49\linewidth]{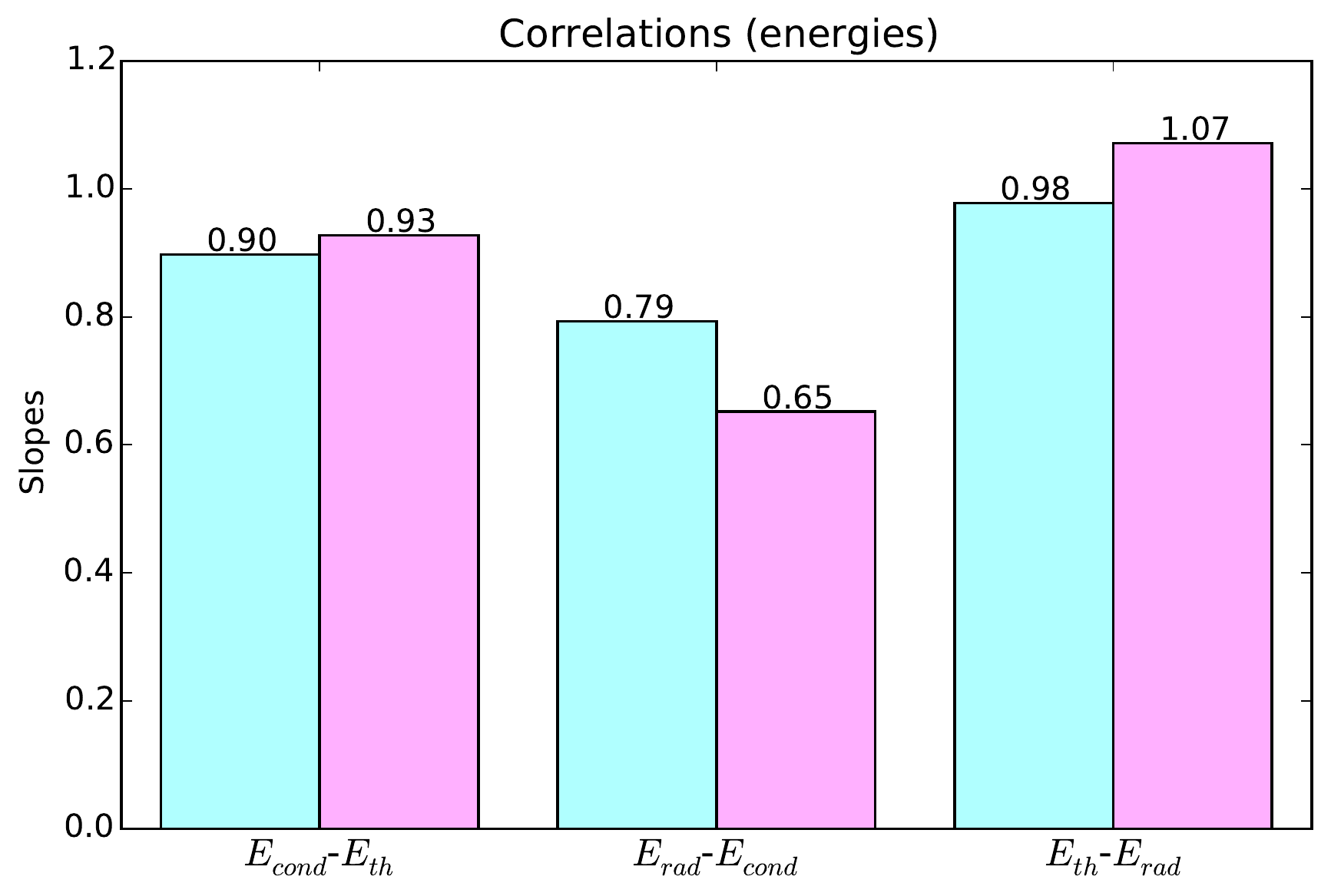}\\
  \includegraphics[width=.49\linewidth]{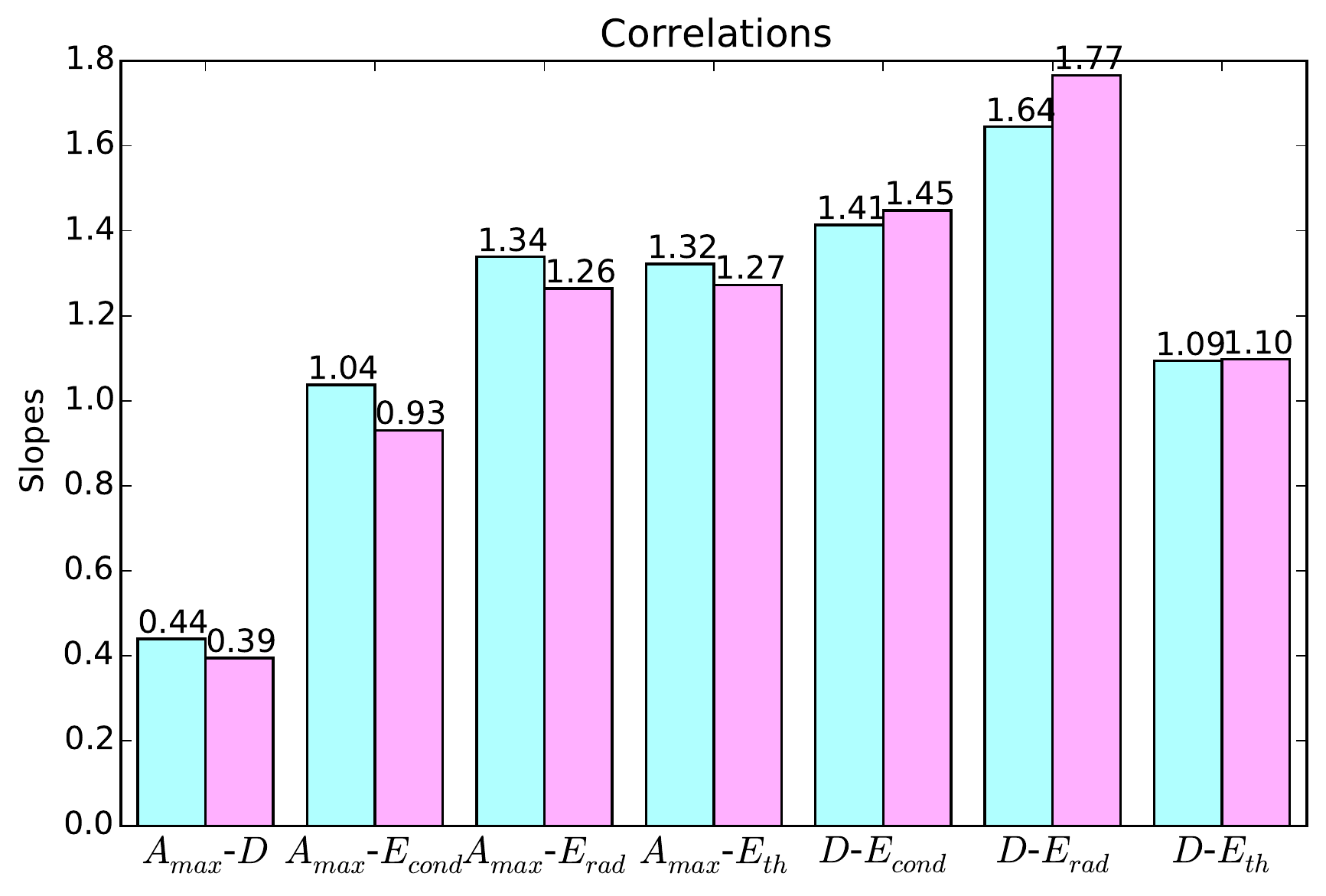}
  \hfill\begin{minipage}[b]{.5\linewidth}
    \caption{Summary of slopes for event parameters distributions and correlations.
    As in other plots, for each distribution or correlation, cyan is for period~1 (first bar of each pair), and magenta is for period~2 (second bar of each pair).}
    \label{fig:slsummary}
  \end{minipage}
\end{figure*}

\subsection{Event parameters statistics}

Power-law slopes for event parameters frequency distributions, as well as for correlations between event parameters, are summarized in Fig.~\ref{fig:slsummary}, which allows the behaviour of slopes mentioned in Sec.~\ref{sec:Results} to be visualized.
For example, we can see what we had already noted for the slopes of event maximum area, duration, and total intensity: they do not vary much between the $17.1$, $19.3$, and $21.1\unit{nm}$ bands; they are $2.23\pm0.04$ for maximum area, $2.47\pm0.12$ for duration, and $1.83\pm0.04$ for total intensity; and they are in general slightly lower in period~2 (corresponding to more large events).
Slopes of event parameters distributions are always larger for the $13.1$ and $33.5\unit{nm}$ bands; this specific behaviour can be due to the fact that these bands include the emission from hotter plasma (see Table~\ref{tab:numbers}).

The slopes for conductive and maximum thermal energy (with no background subtraction) are almost identical, and are lower for radiative energy.
Slopes of background-subtracted energy distributions are systematically lower than for the energies with no background subtraction.
All these slopes are lower than 2, meaning that the total energy in one decade of the less energetic events is less than the total energy in one decade of the most energetic events (although the difference can be small if the power-law index is close enough to the critical value of 2).
The slopes for thermal energy can be compared for example to the slope 1.8 in the compilation of EUV, SXR, and HXR event energies shown in Fig.~10 of \citet{Aschwanden_00}, and to the value 1.66 obtained from AIA for large flares by \citet{aschwanden13b}.
The total powers corresponding to these event energy distributions are discussed more in detail in the next section.

The values in Fig.~\ref{fig:slsummary} allow for an easy comparison with event distribution parameters (typically power-law slopes) from previous works.
For event maximum area distributions, our power-law slopes are higher than the value $2.0$ found in EIT $19.5\unit{nm}$ events by \citet{Berghmans_98}, which is consistent with what \citet{aschwanden13a} find using AIA to analyse large flares, or using a fractal-diffusive avalanche model.
Our slopes for duration distributions are also higher than the value $2.1$ found by \citet{Berghmans_98}.

The correlations between event area and duration (shown also Fig.~\ref{fig:cad}) give an indication on the suitability of the observing cadence used in this study (which determines the shortest detected events) given the instrument spatial resolution and sensitivity (which determine the smallest detected events).
Indeed, we can consider that the shortest detected events should have a duration matching the area of the smallest detected events, according to the fitted correlation; this is the case here.
The same exercise can be done with future instruments, like the Extreme Ultraviolet Imager (EUI) on Solar Orbiter: at the EUI/High Resolution Imager (HRI) pixel size of $0.5\unit{arcsec}$ at a perihelion distance of $0.28\unit{AU}$, the smallest detected events can be expected to be 10 times smaller in area.
Assuming that the correlation between area and duration that we have found here with SDO/AIA can be extrapolated to smaller scales, this means that detecting most of the smallest events at the HRI resolution requires a cadence of about $120/10^{0.4} \approx 45\unit{s}$.

\subsection{Total energy in detected events and contribution to coronal heating}
\label{sec:Total_energy}

\begin{table*}
  \caption{Average power in events, corresponding to the different energies considered in this paper, per unit area of events or per unit area of solar surface. The fraction of the region of interest covered by events, as well as the average total radiated power (in and out of events) per unit area as derived from our measurements are also shown. All powers per unit area are in $\unit{W\cdot m^{-2}}$.}
  \label{tab:avpow}
  \begin{tabular}{lrrrrrrrr}
    \hline\hline
                   & \multicolumn{3}{c}{Power / event area} & \multicolumn{3}{c}{Power / Sun area} & Area fraction & Total radiated power \\
                   & $P_{th,e}$ & $P_{rad,e}$ & $P_{cond,e}$      & $P_{th,S}$ & $P_{rad,S}$ & $P_{cond,S}$   & in events & per unit area \\ \hline
    Obs.\ period~1 & 226 & 88 & 100 & 4.3 & 1.7 & 1.9 & 1.9\% & 61 \\
    Obs.\ period~2 & 777 & 444 & 324 & 15.9 & 9.1 & 6.7 & 2.1\% & 264 \\ \hline
  \end{tabular}

\end{table*}

For any of the energies that we consider (thermal, radiated, or conducted), the average power associated to all events in the region of interest is the sum of energies of all events, divided by the observation duration.
This power can then be further divided by the average area of all events in the field of view, giving an average power per unit area of events (e.g., $P_{th,e}$ for the power associated to thermal energy).

The resulting powers per unit area are shown in Table~\ref{tab:avpow}.
Each of the average thermal, radiative, and conductive power in events, per unit area of events, is about $100$ to $200\unit{W\cdot m^{-2}}$ in period~1 (quiet Sun) and $300$ to $800\unit{W\cdot m^{-2}}$ in period~2 (with a mix of quiet Sun and some active regions).
We note that, in each period, conductive and radiative energies are of the same order of magnitude, as expected from models \citep[e.g.][]{cargill12b}.

In period~1, these powers of energy variation (for thermal energy) or losses (for radiation and conduction) are of the same order of magnitude as the powers expected from the radiative and conductive losses according to \citet{Withbroe_77} in the quiet Sun.
In period~2, these powers are lower than would be expected given the structures visible in the field of view, but the comparison is difficult to do because of the large variations existing between the energy losses of different active regions.

However, if we sum the power in events and divide it by the area of the whole region of interest, we obtain an average power in events per unit area of the whole Sun (e.g., $P_{th,S}$ for thermal energy).
As events cover about $2\%$ of the field of view on average for both observing periods, these average powers are about $50$ times smaller than the ones just discussed above, as shown in Table~\ref{tab:avpow}.
They can also be recovered from integrals of $E \, f(E)$, where $f(E)$ is the frequency distribution of event energy $E$ (e.g., $\Eth$) as shown in Fig.~\ref{fig:disten}.

\newcommand*\emin{\ensuremath{E_\text{min}}}
\newcommand*\emax{\ensuremath{E_\text{max}}}

The resulting powers, which represent the contribution of the detected heating events to the heating of the whole corona, are between $2$ and $4\unit{W\cdot m^{-2}}$ in period~1 and between $7$ and $16\unit{W\cdot m^2}$ in period~2.
These values may seem low, but they are compatible with what can be derived from previous works on energy distributions of heating events, as can be seen from simple computations made on energy frequency distributions modelled by a power-law in the form
\begin{equation}
  f(E) = f_0(E/E_0)^{-\alpha}
\end{equation}
between $\emin$ and $\emax$ (please note that $f_0$ and $E_0$ are not independent parameters).
Such a power-law for energy distribution yields a total power per unit area on the Sun of \citep{Hudson_91,berghmans02p}
\begin{equation}
  P_S=\begin{cases}
    \frac{f_0 E_0^2}{2-\alpha} \left((\emax/E_0)^{2-\alpha} - (\emin/E_0)^{2-\alpha}\right) & \text{for~} \alpha\neq2\\
    f_0 E_0^2 \ln (\emax / \emin) & \text{for~} \alpha=2
  \end{cases}
\end{equation}

For example, if we model our thermal energy frequency distribution in period~1 by a power-law with slope $\alpha=1.9$ between $\emin=2\times10^{17}$ and $\emax=2\times10^{21}\unit{J}$ and a frequency $f_0=10^{-40.3}\unit{s^{-1}\cdot m^{-2}\cdot J^{-1}}$ at the chosen reference energy $E_0=10^{20}\unit{J}$, we obtain a total power (per unit area of the Sun) of about $4.1\unit{W\cdot m^{-2}}$, and so we approximately recover the value $4.3\unit{W\cdot m^{-2}}$ obtained in period~1 by summing all event energies and then dividing them by the observation duration and by the region of interest area.
On the other hand, the parameters of the power-law corresponding to the compilation of frequency distributions made by \citet{Aschwanden_00} are $\alpha=1.8$ and $f_0=10^{-51.5}\unit{s^{-1}\cdot cm^{-2}\cdot erg^{-1}}  =10^{-40.5}\unit{s^{-1}\cdot m^{-2}\cdot J^{-1}}$ at $E_0=10^{27}\unit{erg}=10^{20}\unit{J}$: compared to our power-law, the frequency $f_0$ and the slope $\alpha$ are lower.
This gives then a total power per unit area of the Sun of $2.1\unit{W\cdot m^{-2}}$, which is lower than our results in period~1 but still of the same order of magnitude.

What would happen if the event energy distribution could be extrapolated to smaller energies, including smaller events than the smallest ones that we can detect, like the picoflares mentioned by \citet{parnell00}?
As the slope is smaller than $2$, even doubling the logarithmic energy range of energies, i.e.\ using $\emin=2 \times 10^{13}\unit{J}$ instead of $2 \times 10^{17}\unit{J}$ (with all other parameters of our model distribution remaining the same: $\emax=2\times 10^{21}\unit{J}$, and $f_0=10^{-40.3}\unit{s^{-1}\cdot m^{-2}\cdot J^{-1}}$ at $E_0=10^{20}\unit{J}$), would less than double the total power per unit area of the Sun, giving $5.7\unit{W\cdot m^{-2}}$.
As noted by \citet{berghmans02p}, even a slope of $2$ or $2.1$ would not give more than a few tens watts per square metre, given the observed values $f_0(E_0)$ and \emax, for realistic values of the minimum event energy \emin.

As a point of comparison, the total radiated power that we can compute from the sum of $P_\text{rad,pC}$ (Eq.~\ref{eq:erad}) over the whole region of interest (inside and outside of events) is $61\unit{W\cdot m^{-2}}$ in period~1 and $264\unit{W\cdot m^{-2}}$ in period~2.
This means that the contribution of radiated power in events to the total radiated power of the corona is only $1.7/61 = 2.8\%$ in period~1 and $9.1/264 = 3.4\%$ in period~2.
Such computations would not be possible for the thermal and conduction energies, which rely on assumptions about event geometry (namely height or loop length), but we expect the relative contributions of events to the total thermal energy variation or conductive losses to be low as well.

This points us to a first reason why the contribution of detected heating events to the total coronal heating is low, in this work and in other previous works: the selection of brightening events, standing out of the background, actually excludes most of the energy in the corona.
Such selection excludes in particular continuous processes, as well as high-frequency heating events (occurring at a cadence higher than the plasma cooling time).
In other words, even when the consequences of heating include light emission that is detected by our instruments, heating cannot always be broken up into discrete events, as already noted following the results obtained with SOHO/EIT and TRACE.
The results of our paper show that this difficulty is not overcome by using SDO/AIA observations, with much better cadence, resolution, sensitivity, and spectal coverage.

A second reason is that we do not compute all terms in the energy equation.
For example, with a 1D hydrodynamic equation in coronal loops such the ones used e.g. in \citet{rosner78} or \citet{klimchuk08a}, the values of the time derivative of thermal energy, of divergence of conductive flux, and of the radiative losses terms are not sufficient to determine the heating power: the convective energy flux, the work of pressure and the work of gravity also have to be taken into account.
Even though not all these energies are important all the time, some of them (especially convective energy) can be of the same order of magnitude as the energies that we have computed.
Besides, magnetic energy (which is of course absent from hydrodynamic models) should also be taken into account \citep[as in][for flares]{aschwanden14a}.

Furthermore, SDO/AIA EUV data, corresponding mainly to emission from plasma at warm to moderately hot temperatures, do not constrain the DEM at all temperatures, and wings of the DEM at high and low temperatures are probably underestimated by our inversions.
As a consequence, the energy from this cool and hot plasma can be underestimated, while \citet{brosius14a} have shown that plasma emitting (although faintly) in \ion{Fe}{xix} has significant thermal energy compared to plasma emitting in \ion{Fe}{xii}.
Much energy can also be radiated in white light: in flares, one hundred times more energy is radiated in white light than in soft X-rays \citep{kretzschmar10a}, and this could remain the case for milliflares and below.

For all these reasons (of which most are inherent to what can be done with the available data), the power computed from brightening events in EUV can be expected to represent a small fraction of the total power of coronal heating.

\section{Conclusion}

In the work presented in this paper, we have performed automated detections of EUV brightenings in SDO/AIA images and computed the associated thermal, radiative and conductive energies.
We have presented distributions of different event parameters, including their energies, as well as correlations between them.
The distributions and correlations can be fit by power-laws, with parameters broadly consistent with previous studies.
However, compared to previous studies, the results presented in this paper have been obtained from a large number of events detected using the high-resolution and high-sensitivity SDO/AIA data.
In particular, the event energies have been obtained from DEM inversions computed using all 6 coronal EUV SDO/AIA channels, offering the best possible estimate of energies on this number of events (given the limited coverage of spectroscopic observations).

The computed energies and their distributions give no clue as to whether these events, including the ones that would be too small to be detected (like picoflares, as discussed in Sec.\ \ref{sec:Total_energy}), have sufficient energy to represent a large fraction of the total heating power in the corona, although this is not completely excluded (the distributions could steepen at smaller energies).
However, we have identified some reasons why the average power in detected events only represents a small fraction of the requirements for coronal heating: the process itself of extracting brightening events from the observed corona, the missing knowledge about all terms of the energy equation, and the looseness of observational constraints on the DEM at low and high temperatures.

Then how can we improve the determination of the contribution of heating events (as opposed to continuous heating) to the total coronal heating?

To have a more complete view of events over the accessible parameters range, we could use longer observing periods, but their maximum duration would be limited to a few days by the solar rotation.
We could also use the full SDO/AIA cadence ($12\unit{s}$) for detections, but then the DEM inversions would have either a lower cadence or a lower accuracy, and then we would have less accurate energies.
If observations at very high cadence, resolution, and sensitivity would be possible, we would also be limited by the confusion created when several events occur in the same pixel (or, in extreme cases, by the Olbers-like paradox pointed out in \citet{Aschwanden_00}).

We could also try to evaluate more terms of the energy equation to be able to compute the heating term.
Spectroscopy can provide line-of-sight velocity (and then part of the kinetic energy), better DEM inversions, independent estimates of temperature and density, and then (via the filling factor) an observational constraint on the event height (which is needed to compute thermal energy).
Local reconstructions of the magnetic field from magnetograms would also  allow estimates of the magnetic energy to be derived, as done for flares by \cite{aschwanden14a}.
With current instruments, this would in principle be possible with a combination of SDO/AIA for UV images, SDO/HMI or Hinode/SOT for magnetograms, and Hinode/EIS for spectrograms, provided sufficient common observational coverage is found, during long durations and with high cadence; this is in particular an issue for spectroscopy.
Thanks to numerical simulations, one can also determine how much of the energy involved in heating can actually be retrieved when deriving the energy from observable quantities.

Solar Orbiter/EUI/HRI will provide higher resolution observations than SDO/AIA, potentially during longer durations thanks to quasi-corotation at perihelion.
Observations from higher latitudes could also be interesting.
But HRI includes only one coronal band ($17.4\unit{nm}$), meaning that DEM inversions would have to be done with Solar Orbiter/SPICE data, with lower cadence.
Furthermore, data production will be limited by the low telemetry: an extrapolation of the correlation between event duration and area implies that a cadence of $45\unit{s}$ would be required to match the HRI resolution, and this, with the full HRI field-of-view, would fill the current EUI telemetry allocation for a 5-month orbit in about 3 days (at a data compression rate of 15).
This means that extensive observations of small EUV brightenings cannot be done with Solar Orbiter.
However, if a mission like the JAXA Solar-C mission proposal becomes a reality, we can expect higher cadence and telemetry, as well as better magnetic field reconstructions (starting at the chromosphere instead of the photosphere).
Such new data are critical to understand the contribution of heating events to the global coronal heating, and, more generally, to help understand the detailed physical mechanisms of coronal heating.

\begin{acknowledgements}
  AIA data are courtesy of NASA/SDO and the AIA science team.
  This work made extensive use of the AIA archive at MEDOC, \url{http://medoc-sdo.ias.u-psud.fr}.
  The authors thank Jim Klimchuk, Jean-François Hochedez, Jean-Claude Vial, Pablo Alingery, Aurélien Canou, and Élie Soubrié for interesting discussions and for their help with this work.
  Financial support from CNES for travel and for building the AIA archive at MEDOC is also acknowledged.
\end{acknowledgements}

\bibliographystyle{aa}
\bibliography{biblio}

\begin{thebibliography}{46}
\expandafter\ifx\csname natexlab\endcsname\relax\def\natexlab#1{#1}\fi

\bibitem[{{Aletti} {et~al.}(2000){Aletti}, {Velli}, {Bocchialini}, {Einaudi},
  {Georgoulis}, \& {Vial}}]{aletti00}
{Aletti}, V., {Velli}, M., {Bocchialini}, K., {et~al.} 2000, \apj, 544, 550

\bibitem[{{Alfv{\' e}n}(1941)}]{alf41}
{Alfv{\' e}n}, H. 1941, Arkiv för Matematik, Astronomi och Fysik, Band 27A, 1

\bibitem[{{Alfv{\' e}n}(1947)}]{alf47}
{Alfv{\' e}n}, H. 1947, \mnras, 107, 211

\bibitem[{{Aschwanden} \& {Shimizu}(2013)}]{aschwanden13b}
{Aschwanden}, M.~J. \& {Shimizu}, T. 2013, \apj, 776, 132

\bibitem[{{Aschwanden} {et~al.}(2000){Aschwanden}, {Tarbell}, {Nightingale},
  {Schrijver}, {Title}, {Kankelborg}, {Martens}, \& {Warren}}]{Aschwanden_00}
{Aschwanden}, M.~J., {Tarbell}, T.~D., {Nightingale}, R.~W., {et~al.} 2000,
  \apj, 535, 1047

\bibitem[{{Aschwanden} {et~al.}(2014){Aschwanden}, {Xu}, \&
  {Jing}}]{aschwanden14a}
{Aschwanden}, M.~J., {Xu}, Y., \& {Jing}, J. 2014, \apj, 797, 50

\bibitem[{{Aschwanden} {et~al.}(2013){Aschwanden}, {Zhang}, \&
  {Liu}}]{aschwanden13a}
{Aschwanden}, M.~J., {Zhang}, J., \& {Liu}, K. 2013, \apj, 775, 23

\bibitem[{{Auch{\`e}re} {et~al.}(2014){Auch{\`e}re}, {Bocchialini}, {Solomon},
  \& {Tison}}]{auchere14a}
{Auch{\`e}re}, F., {Bocchialini}, K., {Solomon}, J., \& {Tison}, E. 2014, \aap,
  563, A8

\bibitem[{{Bak} {et~al.}(1988){Bak}, {Tang}, \& {Wiesenfeld}}]{bak88}
{Bak}, P., {Tang}, C., \& {Wiesenfeld}, K. 1988, \pra, 38, 364

\bibitem[{{Berghmans}(2002)}]{berghmans02p}
{Berghmans}, D. 2002, in ESA Special Publication, Vol. 506, Solar Variability:
  From Core to Outer Frontiers, ed. A.~{Wilson}, 501--508

\bibitem[{{Berghmans} {et~al.}(1998){Berghmans}, {Clette}, \&
  {Moses}}]{Berghmans_98}
{Berghmans}, D., {Clette}, F., \& {Moses}, D. 1998, \aap, 336, 1039

\bibitem[{{Brosius} {et~al.}(2014){Brosius}, {Daw}, \& {Rabin}}]{brosius14a}
{Brosius}, J.~W., {Daw}, A.~N., \& {Rabin}, D.~M. 2014, \apj, 790, 112

\bibitem[{{Buchlin} {et~al.}(2003){Buchlin}, {Aletti}, {Galtier}, {Velli},
  {Einaudi}, \& {Vial}}]{buchlin03}
{Buchlin}, E., {Aletti}, V., {Galtier}, S., {et~al.} 2003, \aap, 406, 1061

\bibitem[{{Buchlin} {et~al.}(2005){Buchlin}, {Galtier}, \& {Velli}}]{buchlin05}
{Buchlin}, E., {Galtier}, S., \& {Velli}, M. 2005, \aap, 436, 355

\bibitem[{{Cargill} {et~al.}(2012){Cargill}, {Bradshaw}, \&
  {Klimchuk}}]{cargill12b}
{Cargill}, P.~J., {Bradshaw}, S.~J., \& {Klimchuk}, J.~A. 2012, \apj, 758, 5

\bibitem[{{Crosby} {et~al.}(1993){Crosby}, {Aschwanden}, \&
  {Dennis}}]{crosby93}
{Crosby}, N.~B., {Aschwanden}, M.~J., \& {Dennis}, B.~R. 1993, \solphys, 143,
  275

\bibitem[{{Dennis}(1985)}]{dennis85}
{Dennis}, B.~R. 1985, \solphys, 100, 465

\bibitem[{{Drake}(1971)}]{Drake_71}
{Drake}, J.~F. 1971, \solphys, 16, 152

\bibitem[{{Edl{\' e}n}(1943)}]{edlen43}
{Edl{\' e}n}, B. 1943, Zeitschrift fur Astrophysics, 22, 30

\bibitem[{{Einaudi} {et~al.}(1996){Einaudi}, {Velli}, {Politano}, \&
  {Pouquet}}]{einaudi96}
{Einaudi}, G., {Velli}, M., {Politano}, H., \& {Pouquet}, A. 1996, \apjl, 457,
  L113

\bibitem[{{Georgoulis} {et~al.}(1998){Georgoulis}, {Velli}, \&
  {Einaudi}}]{georgoulis98}
{Georgoulis}, M.~K., {Velli}, M., \& {Einaudi}, G. 1998, \apj, 497, 957

\bibitem[{{Grotrian}(1939)}]{grotrian39}
{Grotrian}, W. 1939, Naturwissenschaften, 27, 214

\bibitem[{{Guennou} {et~al.}(2012){Guennou}, {Auch{\`e}re}, {Soubri{\'e}},
  {Bocchialini}, {Parenti}, \& {Barbey}}]{guennou12b}
{Guennou}, C., {Auch{\`e}re}, F., {Soubri{\'e}}, E., {et~al.} 2012, \apjs, 203,
  26

\bibitem[{{Hannah} {et~al.}(2008){Hannah}, {Christe}, {Krucker}, {Hurford},
  {Hudson}, \& {Lin}}]{hannah08b}
{Hannah}, I.~G., {Christe}, S., {Krucker}, S., {et~al.} 2008, \apj, 677, 704

\bibitem[{{Hortin}(2003)}]{hortin}
{Hortin}, T. 2003, PhD thesis, Univ. Paris-Sud, Orsay, France

\bibitem[{{Hudson}(1991)}]{Hudson_91}
{Hudson}, H.~S. 1991, \solphys, 133, 357

\bibitem[{{Klimchuk} {et~al.}(2008){Klimchuk}, {Patsourakos}, \&
  {Cargill}}]{klimchuk08a}
{Klimchuk}, J.~A., {Patsourakos}, S., \& {Cargill}, P.~J. 2008, \apj, 682, 1351

\bibitem[{{Koutchmy} {et~al.}(1997){Koutchmy}, {Hara}, {Suematsu}, \&
  {Reardon}}]{koutchmy97a}
{Koutchmy}, S., {Hara}, H., {Suematsu}, Y., \& {Reardon}, K. 1997, \aap, 320,
  L33

\bibitem[{{Kretzschmar} {et~al.}(2010){Kretzschmar}, {de Wit}, {Schmutz},
  {Mekaoui}, {Hochedez}, \& {Dewitte}}]{kretzschmar10a}
{Kretzschmar}, M., {de Wit}, T.~D., {Schmutz}, W., {et~al.} 2010, Nature
  Physics, 6, 690

\bibitem[{{Krucker} \& {Benz}(1998)}]{Krucker_98}
{Krucker}, S. \& {Benz}, A.~O. 1998, \apjl, 501, L213

\bibitem[{{Lemen} {et~al.}(2012){Lemen}, {Title}, {Akin}, {Boerner}, {Chou},
  {Drake}, {Duncan}, {Edwards}, {Friedlaender}, {Heyman}, {Hurlburt}, {Katz},
  {Kushner}, {Levay}, {Lindgren}, {Mathur}, {McFeaters}, {Mitchell}, {Rehse},
  {Schrijver}, {Springer}, {Stern}, {Tarbell}, {Wuelser}, {Wolfson}, {Yanari},
  {Bookbinder}, {Cheimets}, {Caldwell}, {Deluca}, {Gates}, {Golub}, {Park},
  {Podgorski}, {Bush}, {Scherrer}, {Gummin}, {Smith}, {Auker}, {Jerram},
  {Pool}, {Soufli}, {Windt}, {Beardsley}, {Clapp}, {Lang}, \&
  {Waltham}}]{lemen12a}
{Lemen}, J., {Title}, A., {Akin}, D., {et~al.} 2012, \solphys, 275, 17

\bibitem[{{Lin} {et~al.}(1984){Lin}, {Schwartz}, {Kane}, {Pelling}, \&
  {Hurley}}]{lin84}
{Lin}, R.~P., {Schwartz}, R.~A., {Kane}, S.~R., {Pelling}, R.~M., \& {Hurley},
  K.~C. 1984, \apj, 283, 421

\bibitem[{{Lu} \& {Hamilton}(1991)}]{luh91}
{Lu}, E.~T. \& {Hamilton}, R.~J. 1991, \apjl, 380, L89

\bibitem[{{Madjarska} \& {Doyle}(2002)}]{Madjarska_02}
{Madjarska}, M.~S. \& {Doyle}, J.~G. 2002, in ESA Special Publication, Vol.
  508, From Solar Min to Max: Half a Solar Cycle with SOHO, ed. A.~{Wilson},
  311--314

\bibitem[{{Madjarska} {et~al.}(2003){Madjarska}, {Doyle}, {Teriaca}, \&
  {Banerjee}}]{Madjarska_03}
{Madjarska}, M.~S., {Doyle}, J.~G., {Teriaca}, L., \& {Banerjee}, D. 2003,
  \aap, 398, 775

\bibitem[{{Parker}(1983)}]{parker83}
{Parker}, E.~N. 1983, \apj, 264, 635

\bibitem[{{Parker}(1988)}]{Parker_88}
{Parker}, E.~N. 1988, \apj, 330, 474

\bibitem[{{Parnell} \& {Jupp}(2000)}]{parnell00}
{Parnell}, C.~E. \& {Jupp}, P.~E. 2000, \apj, 529, 554

\bibitem[{{Pesnell} {et~al.}(2012){Pesnell}, {Thompson}, \&
  {Chamberlin}}]{pesnell12a}
{Pesnell}, W., {Thompson}, B., \& {Chamberlin}, P. 2012, \solphys, 275, 3

\bibitem[{{Rosner} {et~al.}(1978){Rosner}, {Tucker}, \& {Vaiana}}]{rosner78}
{Rosner}, R., {Tucker}, W.~H., \& {Vaiana}, G.~S. 1978, \apj, 220, 643

\bibitem[{{Sornette}(2000)}]{sorn}
{Sornette}, D. 2000, Critical Phenomena in Natural Sciences: Chaos, Fractals,
  Selforganization and Disorder (Springer-Verlag)

\bibitem[{{Spitzer} \& {H{\"a}rm}(1953)}]{spitzerharm53}
{Spitzer}, L. \& {H{\"a}rm}, R. 1953, Phys. Rev., 89, 977

\bibitem[{{Thompson}(2006)}]{thompson06}
{Thompson}, W.~T. 2006, \aap, 449, 791

\bibitem[{{van Ballegooijen}(1986)}]{vanballegooijen86}
{van Ballegooijen}, A.~A. 1986, \apj, 311, 1001

\bibitem[{{Winebarger} {et~al.}(2013){Winebarger}, {Walsh}, {Moore}, {De
  Pontieu}, {Hansteen}, {Cirtain}, {Golub}, {Kobayashi}, {Korreck}, {DeForest},
  {Weber}, {Title}, \& {Kuzin}}]{winebarger13a}
{Winebarger}, A.~R., {Walsh}, R.~W., {Moore}, R., {et~al.} 2013, \apj, 771, 21

\bibitem[{{Withbroe} \& {Noyes}(1977)}]{Withbroe_77}
{Withbroe}, G.~L. \& {Noyes}, R.~W. 1977, \araa, 15, 363

\end{thebibliography}

\end{document}